\documentclass[a4paper,11pt,dvipsnames]{aa}  
\usepackage{graphicx}
\usepackage{comment}
\usepackage{textcomp}
\usepackage{txfonts}
\usepackage{multicol}
\usepackage{siunitx}
\usepackage{booktabs}
\usepackage{array}
\usepackage{dcolumn}
\usepackage{color}
\usepackage{xcolor}
\usepackage{eufrak}
\usepackage{booktabs}
\usepackage{adjustbox}					
\usepackage[mathscr]{euscript}
\definecolor{v}{rgb}{0.54, 0.17, 0.89} 
\usepackage{soul}
\usepackage{lscape}
\usepackage{adjustbox}
\usepackage{tabularx}
\usepackage{float}
\usepackage{comment}
\usepackage{amsmath}
\usepackage{xcolor}
\usepackage{ulem}
\usepackage{lipsum}
\usepackage{rotating}
\usepackage{lscape}
\usepackage{cellspace}
\usepackage{tablefootnote}
\usepackage{caption}
\usepackage[skip=0.5ex]{subcaption}

\usepackage{epigraph}
\begin{document}

   \title{Black Hole masses for 14 gravitational lensed quasars}

   \author{A. Melo
          \inst{1,2,3}
          \and
          V. Motta\inst{3}
          \and
          J. Mejía-Restrepo\inst{4}
          \and
          R. J. Assef\inst{5}
          \and   
          N. Godoy\inst{3, 6, 7}
          \and
          E. Mediavilla\inst{8,9}
          \and
          E. Falco\inst{10}
          \and
          C. S. Kochanek\inst{11, 12}
          \and
          F. Ávila-Vera\inst{3}
          \and
          R. Jerez\inst{3}.
          }

   \institute{Max-Planck-Institut für Astrophysik, Karl-Schwarzschild-Str. 1, 85748 Garching, Germany\\
              \email{amelo@mpa-garching.mpg.de}
         \and
    Technical University of Munich, TUM School of Natural Sciences, Department of Physics, James-Franck-Straße 1, 85748 Garching, Germany
         \and
            Instituto de Físisca y Astronomía, Facultad de Ciencias, Universidad de Valparaíso, Av. Gran Bretaña 1111, Valparaíso, Chile.
        \and
        EPAM Systems,41 University Drive, Suite 202 Newtown, PA 18940, USA.
        \and
    Instituto de Estudios Astrofísicos, Facultad de Ingeniería y Ciencias, Universidad Diego Portales, Av. Ejército Libertador 441, Santiago 8320000, Chile.
        \and
        N\'ucleo Milenio de Formaci\'on Planetaria (NPF), 2360102, Chile.
        \and
        Aix Marseille Univ, CNRS, CNES, LAM, Marseille, France.
        \and
    Instituto de Astrofísica de Canarias, Vía Láctea S/N, La Laguna 38200, Tenerife, Spain.
    \and
    Departamento de Astrofísica, Universidad de la Laguna, La Laguna 38200, Tenerife, Spain.
    \and
    Harvard-Smithsonian Center for Astrophysics, 60 Garden Street, Cambridge, MA, 02138, USA.
    \and
    Department of Astronomy, The Ohio State University, 140 West 18th Avenue, Columbus, OH, 43210, USA.
    \and
    Center for Cosmology and Astroparticle Physics, The Ohio State University, 191 W. Woodruff Avenue, Columbus, OH, 43210, USA.
             }

   \date{Received -- --, ----; accepted -- --, ----}

 
  \abstract
   {}
   {We estimate black hole masses (M$_{\rm BH}$) for 14 gravitationally lensed quasars using the Balmer lines along with estimates based on Mg{\footnotesize{II}} and C{\footnotesize{IV}} emission lines for four and two of them, respectively. We compare with results obtained for other lensed quasars.}
   {We use spectroscopic data from the Large Binocular Telescope (LBT), Magellan and the Very Large Telescope (VLT) to measure the FWHM of the broad emission lines. Combined with the bolometric luminosity measured from the spectra energy distribution, we estimate M$_{\rm BH}$ including uncertainties from microlensing and variability.
   }
   {We obtain M$_{\rm BH}$ using the single-epoch method from the H$\alpha$ and/or H$\beta$ broad emission lines for 14 lensed quasars, including the first estimates for QJ0158$-$4325, HE0512$-$3329 and WFI2026$-$4536. The masses are typical of non-lensed quasars of similar luminosity, and the implied Eddington ratios are typical. We have increased the sample of lenses with estimates of M$_{\rm BH}$ by 60$\%$.
   }
   {}


   \maketitle
%

\section{Introduction}
 
Supermassive black holes (SMBHs) are thought to be a key ingredient in galaxy formation and evolution, particularly since the discovery that the central SMBH mass (M$_{\rm{BH}}$) has a tight correlation with the stellar luminosity and velocity dispersion (\citealt{1995Kormendy,2000Ferrarese, 2002Tremaine,2003Marconi,2013kormendyandho, 2019GReGr..51...65Z}) of the spheroidal components of their host galaxies. To understand this link, we need to study the evolution of the SMBH, their hosts and their environments, particularly during the phases with significant accretion rates when the active galactic nucleus (AGN) is releasing large amounts of energy (see, e.g., \citealt{2005DiMatteo,2006Croton,2008Hopkins}). Reliably measuring M$_{\rm{BH}}$ is fundamental to understanding this connection.\\
In the unified model of AGN (\citealt{1993ARA&A..31..473A,1995Urry}), the accretion disk continuum emission illuminates nearby gas to produce the broad emission lines (BELs) in the spectra. Continuum variability drives a delayed change in the BEL fluxes and line profiles. Reverberation mapping (RM, \citealt{1993PASP..105..247P,1997ASSL..218...85N} and therein) measures this delay to determine the size of the BEL region (\citealt{1999Wandel,2000kaspi,2004Peterson,2009Bentz}), which can then be used to estimate M$_{\rm{BH}}$ given the line widths and local calibrations. Even locally, RM is challenging because it requires repeated spectroscopic observations over months (\citealt{2004Peterson,2009Bentz,2015Barth,2017Grier,2019Grier,2016Du,2018Lira}), and the required monitoring periods increase for more luminous quasars or, due to time dilation, higher redshift quasars (\citealt{2018Lira}). Initially, RM studies were largely limited to individual studies of
local, lower luminosity quasars, but the samples have recently expanded to higher luminosities and redshifts by using multi-fiber spectrographs to monitor hundreds of AGN simultaneously (\citealt{2023Malik,2023Shen,2023Yu}). Nonetheless, current RM samples have only $\sim 10^2$ AGN, and it will be a long process to reach $\sim 10^3$ AGN. 
Fortunately, RM revealed  a correlation between the BLR distance from the BH and the optical continuum luminosity, known as the size-luminosity (R-L) relation (\citealt{2005kaspi,2006Bentz,2011Zu}). This relationship combined with the virial theorem allows us to estimate M$_{\rm{BH}}$ using a single spectrum, a procedure known as the single-epoch (SE) method (e.g. \citealt{2004MclureDunlop,2006Vestergaard,2011shen,2012shenliu}). The SE method was developed and calibrated using the H$\beta$ width (e.g. \citealt{2004ApJvestergaard,2011Xiao,2012shenliu}).\\
For higher redshift systems (z $>$ 0.9), H$\beta$ is shifted into the Near Infrared (NIR), making it difficult to observe large samples from the ground due to the bright sky emission. One solution is to instead use the Mg{\footnotesize{II}} or C{\footnotesize{IV}} lines (\citealt{2002Mclure,2002Vestergaard}) to study z $>$ 0.9 systems in the optical (e.g. \citealt{2008Mcgill,2013Park,2015Park,2017Coatman,2018Woo}). However, this approach present several drawbacks: 1) these UV lines lack a local calibration because they cannot be observed from the ground, 2) their indirect calibrations are restricted to high-luminosity objects (\citealt{2016Mejiarestrepo}), 3) Mg{\footnotesize{II}} may have a small but significant dependence on the Eddington ratio of the AGN and might not be reliable in objects with FWHM(Mg{\footnotesize{II}})$\geqslant$ 6000 km/s (\citealt{2013marziani}), and 4) there are concerns regarding C{\footnotesize{IV}} because its width could be affected by winds of ejected disk material (\citealt{2011assef,2016Coatman,2018mejia-restrepob}) and microlensing in the case of lensed QSOs (\citealt{2018fian}). 
The C{\footnotesize{IV}} emission line is more asymmetric than the Balmer lines and Mg{\footnotesize{II}}, and its width is not well correlated with those of H$\beta$ and Mg{\footnotesize{II}} (e.g., \citealt{2005Baskin,2008Shen}), but early studies showed a strong correlation between the width of H$\alpha$, H$\beta$ and Mg{\footnotesize{II}} (see \citealt{2005Greene,2008Shen,2009Wang,2012shenliu}). Hence, it is reasonable to argue that the virial mass estimator based on the Balmer lines is the most reliable one. The H$\beta$ emission line is typically preferred (due to its wavelength and lack of blended emission lines), and H$\alpha$ is also known to work well (\citealt{2005Greene,2007Netzer,2011Xiao}).\\
Many studies have estimated M$_{\rm BH}$ using the SE method for large samples of quasars (e.g. \citealt{2002Mclure,2004MclureDunlop,2006Vestergaard,2013shen,2014Peterson,2016Mejiarestrepo,2019shen}), and it has also been used to estimate M$_{\rm BH}$ for samples of lensed AGNs. Gravitational lenses allow us to investigate the inner structure of lensed quasars (see, e.g., \citealt{2004Kochanek,2010morgan}). \citet{2006Peng} was the first to estimate the M$_{BH}$ of 31 gravitationally lensed AGNs. They applied the virial technique using the C{\footnotesize{IV}} (22 systems), Mg{\footnotesize{II}} (19 systems) and H$\beta$ (two systems) emission line widths and the continuum luminosities $\lambda$L$_{\lambda}$ at 1300, 3000 and 5100~\AA, respectively. Seven of the systems have estimates obtained from two different emission lines. 
\citet{2010greene} obtained M$_{\rm BH}$ for 11 systems using H$\alpha$ and H$\beta$ (nine have both). Their goal was to search for systematic biases in the \citet{2006Peng} M$_{\rm BH}$ estimates due to the use of the C{\footnotesize{IV}} emission line. Even though the masses presented by \citet{2010greene} are more robust (they used spectra with higher S/N), they conclude that there is no evidence for a systematic bias between the lines used by \citet{2006Peng} and the Balmer lines, despite the large scatter. \citealt{2011assef} searched for possible biases between M$_{\rm BH}$ estimates based on the H$\alpha$, H$\beta$ and C{\footnotesize{IV}} broad emission lines, improving the sample with new observations and adding missing luminosity estimates at $\lambda=5100$~\AA~. They selected 12 lensed quasars from the CfA-Arizona Space TElescope LEns Survey (CASTLES\footnote{\url{https://lweb.cfa.harvard.edu/castles/}}, \citealt{2001castle}) with high quality C{\footnotesize{IV}} spectra and published NIR spectra of the Balmer lines. The FWHM were obtained using broad and narrow Gaussian components and the continuum luminosity at 5100~\AA~ was estimated using the AGN spectral energy distribution (SED) template of \citet{2010Assef}. They conclude that the M$_{\rm BH}$ inferred from C{\footnotesize{IV}} using the line dispersion ($\sigma_l$) shows a systematic offset with respect to the estimate using the FWHM. However, \citet{2011assef} compared the M$_{\rm BH}$ estimated using C{\footnotesize{IV}} and the Balmer lines and found no significant offset.
\citet{2012Sluse}, in a study of microlensing in a sample of 17 lensed quasars, obtained M$_{\rm BH}$ using the C{\footnotesize{IV}} (5 systems), Mg{\footnotesize{II}} (12 systems) and H$\beta$ emission lines (2 systems), where two objects have estimates from two different emission lines and four had published values from \citet{2006Peng} and \citet{2011assef}. 

There have been no new M$_{\rm BH}$ estimates for lensed quasars in the last
decade. In general, recent publications refer to the M$_{\rm BH}$ mentioned above (e.g.  \citealt{2017dingII,2020Guerras,2021Ding,2021Hutsemekers}), and only 14 of the 222\footnote{Gravitationally Lensed Quasar Database, GQL \url{https://research.ast.cam.ac.uk/lensedquasars/index.html}} known lensed quasars have M$_{\rm BH}$ measurements based on the H$\alpha$ and/or H$\beta$ lines.
In this work, we increase the sample of Balmer lines M$_{\rm BH}$ estimates for lensed AGNs from 14 to 23 sources. Even though the majority of the objects in our sample (with the exception of WFI2026$-$4536 and HE0512$-$3329) have BH mass estimates (\citealt{2006Peng,2011assef,2012Sluse,2017dingII}), only two of them (SDSS1138+0314 and HE1104$-$1805) were obtained using  H$\alpha$ or H$\beta$. Most are based on the C{\footnotesize{IV}} and/or Mg{\footnotesize{II}} broad emission lines. We also include three quasars with no previous $M_{\rm BH}$ estimates.

This paper is structured as follows. In Sect.~\ref{sec:Obs} we present the systems and data reduction for the three different instruments used in this work (VLT/X-shooter, LBT/LUCI and Magellan/MMIRS). Section~\ref{sec:smbh} describes the method for obtaining M$_{BH}$ and the factors that could contribute to its uncertainties. Our results are presented in section~\ref{Results}, analyzing the systems and comparing with the large samples of non-lensed AGNs. Finally, our conclusions are presented in section~\ref{Conclusion}. Throughout the text we assume a $\Lambda$CDM cosmology with $\Omega_{\Lambda}=0.7$, $\Omega_{M}=0.3$ and $H_{O}=70~ \rm km s^{-1}~Mpc^{-1}$.

\section{Observations and Data Reduction} \label{sec:Obs}

We present observations for three systems with the X-shooter instrument (\citealt{2011vernet}) and one observation with the FOcal Reducer/low dispersion Spectrograph 2 (FORS2, \citealt{fors}) at the Very Large Telescope (VLT). In addition, we include 21 spectroscopic observations taken in 2012 for 14 lensed quasars with the Large Binocular Telescope (LBT) and the LUCI spectograph (\citealt{2003SPIE.4841..962S}) or the Magellan telescope and the MMT and Magellan Infrared Spectrograph (MMIRS; \citealt{2012PASP..124.1318M}). Table~\ref{table:observation} summarizes the main observational characteristics for the observing runs, the image(s) observed for each lensed quasar and the orientation of the slit. Data reduction for each instrument is described below.

\begin{table*}
\caption{Observations}             
\label{table:observation}      
\centering          
\begin{tabular}{ l c c c c c c }     
\hline\hline       

Object & Date & Position angle & Exp. time & Image(s) & Filter & Inst.\\
& (dd-mm-YYYY) & (\textdegree) & (s) &  & &  \\ 
\hline                    
HE0047$-$1756 & 25-Nov-2012 & 354.121 & 120 & A  & HKspec & LUCIFER \\
HE0047$-$1756 & 25-Nov-2012 & 354.121 & 120 & A  &  J & LUCIFER \\
HE0435$-$1223 & 27-Nov-2012 & 303.674 & 120 & A  & HKspec & LUCIFER \\
HE0512$-$3329 & 06-Apr-2012 & 85.409 & 180-300 & A-B & HK & MMIRS \\
SDSS0924+0219 & 24-Nov-2012 & 361.326 & 120 & A  &  HKspec & LUCIFER \\
SDSS0924+0219 & 24-Nov-2012 & 361.326 & 600 & A   & J & LUCIFER \\
Q1017$-$207 & & & & A-B & HK & MMIRS \\
HE1104$-$1805 & 07-Apr-2012 & 131.361 & 180-300 & A-B  & HK & MMIRS \\
SDSS1138$+$0314 & 06-Apr-2012 & 93.836 & 180-300 & A-B  & HK & MMIRS \\
SDSSJ1335$+$0118 & 07-Apr-2012 & 90.246  & 180-300 & A-B  & HK & MMIRS \\
WFI2026$-$4536 & 06-Nov-2012 & 19.798 & 180 & A-B  & HK & MMIRS \\
WFI2033$-$4723 & 06-Apr-2012 & 59.148  & 180-300 & C-A2  & HK & MMIRS \\
HE2149$-$2745 & 06-Apr-2012 & 38.578 & 180-300 & A-B  & HK & MMIRS \\
QJ0158$-$4325 & 21/22-Aug-2019 & 70.98 & 600x8 & A-B & UVB, VIS and NIR & X-shooter\\
QJ0158$-$4325 & 19/20-Sep-2019 & 70.98 & 600x8 & A-B & UVB, VIS and NIR & X-shooter\\
SDSS1226$-$0006 & 6/7-Mar-2013 & 87.5 & 600x4 & A-B & UVB, VIS and NIR (JH) & X-shooter\\
SDSS1226$-$0006 & 10/13-Feb/2010 & -91.89 & 2800x4 & A-B & VIS & FORS2\\
LBQS1333+0113 & 27/28-Feb-2020 & 138.439 & 600x8 & A-B & UVB, VIS and NIR & X-shooter\\
LBQS1333+0113 & 28/29-Feb-2020 & 138.439 & 600x8 & A-B & UVB, VIS and NIR & X-shooter\\
Q1355$-$2257 & 28/29-Feb-2020 & -106.467 & 600x8 & A-B & UVB, VIS and NIR & X-shooter\\
Q1355$-$2257 & 6/7-Apr-2021 & -106.467 & 600x8 & A-B & UVB, VIS and NIR & X-shooter\\
\hline                  
\end{tabular}
\end{table*}

\subsection{X-shooter}

LBQS1333+0113, QJ0158$-$4325 and Q1355$-$2257 were observed with X-shooter between August of 2019 and April of 2021 (ESO proposal ID $\rm 103.B-0566(A)$; PI: A. Melo).  We used two Observing Blocks (OBs) for each system with a slit width of  $1\farcs0 \times 11\arcsec$ for the UVB band (resolution of R = 5400) and  $1\farcs2 \times 11\arcsec$ for VIS and NIR arm (R = 6500 and 4300 respectively). In the first OB, four exposures were taken in the NIR arm (600s each) and two exposures in the VIS and UVB arm (600s each), with a nodding of 3$\arcsec$ per frame and a readout mode (UVB and VIS) of 100k/1pt/hg. The second OB had the same configuration as the first one, but the NIR data was taken with two exposures instead of four. The slit was centered on the brightest image of the lensed quasar and the position angle was chosen to include the second brightest image. We used the atmospheric dispersion corrector (ADC) to correct for differential atmospheric refraction. SDSS1226$-$0006 was observed in 2013, with slit width of $1\farcs6 \times 11\arcsec$ for the UVB band,  $1\farcs5 \times 11\arcsec$ for VIS and $0\farcs9 \times 11\arcsec$ NIR arm.

The data were reduced using the ESO pipeline \texttt{EsoReflex} (\citealt{2013esoreflex}) along with Principal Component Analysis (PCA; \citealt{1964DeemingPCA,1981BujarrabalPCA,1999FrancisPCA}) for the sky emission subtraction. We briefly summarize the steps here (more details can be found in \citealt{2021Melo}). First, X-shooter pipeline version 3.5.0 of \texttt{EsoReflex} was used to reduce each individual OB (flat field, dark current, wavelength calibration, among others) without correction for nodding and without subtracting the sky background. We used PCA for the sky emission correction in the NIR on each individual frame. First, we masked outliers (such as bad pixels) using $\sigma -$clipping and replace them with a value from a bicubic interpolation of the surrounding pixels. We calculated a sky median as a function of wavelength, subtract it from each frame and collapse the two dimensional (2D) spectra along the wavelength axis to select an uncontaminated spatial region for the sky emission. We chose the PCA-basis as the region of threshold equal to 3 of the median above the background (see Fig. 2 of \citealt{2021Melo}). Finally, we constructed a model of the sky emission in the selected spatial region as our PCA eigenvector basis and subtracted it from the frame.

Flux calibration is done by using equation 3 of the X-shooter Pipeline User Manual\footnote{\url{https://ftp.eso.org/pub/dfs/pipelines/instruments/xshooter/xshoo-pipeline-manual-3.5.3.pdf}} with the response curve from the X-shooter pipeline based on a standard star observed the same night as the target.

We used \texttt{molecfit} (\citealt{2015A&A...576A..77S,2015Kauschmolecfit}) for the telluric correction of each spectrum and employed the best fit to each spectrum row by row. Finally, the spectra were median combined using the parameters from the header for the stacking. The uncertainties were estimated as the median absolute deviation.

For the VIS and UVB reduction, we used a median of each sky region as the model of the sky brightness, but otherwise followed the same steps used for the NIR.

\subsection{LUCI}

The systems HE0047$-$1756, HE0435$-$1223, SDSS0924+0219, and Q1017$-$207 were observed (November 24 to 27 of 2012) in the longslit mode using the gratings 200$\_$H$+$K (with a resolving power of 1881 at H and 2573 at K) and 210$\_$zJHK (a resolving power of 6877, 8460, 7838 and 6687 at z, J, H and K respectively) with a 0\farcs5 wide slit. The N1.8 camera was used with a pixel scale of 0\farcs25. The estimated seeing was $\sim 0\farcs8$.

Data reduction was performed using IRAF packages along with IDL task {\tt xtellcor$\_$general} from \cite{2003PASP..115..389V} for the telluric absorption correction. The detailed reduction is described in \citealt{2011assef}, but we present a summary of the steps here. For each exposure, a two-dimensional wavelength calibration was performed using the sky emission lines, and a combined median sky frame was built. This sky frame was used to remove the sky before extracting the spectra. The telluric absorption correction was made using {\tt xtellcor$\_$general}.

\subsection{MMIRS}

Seven lensed systems were observed using MMIRS
on 2012 April 6 and 7 using the long-slit data spanning H/K bands (1.25-2.4 $\mu$m). Two images of the lensed quasar were positioned in a slit of 0\farcs8~ wide with a pixel scale of 0\farcs2012\footnote{\url{https://lweb.cfa.harvard.edu/mmti/mmirs/instrstats.html}}. The spectra were taken with nodding to control for the sky background.

Data reductions were carried out with the instrument pipeline \citep{mmirs_pipe} and IRAF\footnote{IRAF is distributed by the National Optical Astronomy Observatory, which is operated by the Association of Universities for Research in Astronomy, Inc., under cooperative agreement with the National Science Foundation.} tasks. The code {\tt mmfixall}, provided by the MMIRS instrument scientific team, was used to collapse the information contained in the multi-extension files. The remaining procedures were performed in IRAF and consisted of dark correction, sky subtraction, 1D spectra extraction, wavelength calibration and telluric correction. The 1D spectra was extracted using the {\tt apall} task with apertures of $\pm~3-4$ pixels. Flux calibration was carried out using {\tt xtellcor$\_$general} for telluric absorption corrections.

\subsection{FORS2}

Only SDSS1226$-$0006 was observed using FORS2 on February of 2010. Data reduction was performed using IRAF and standard procedure consisting of bias subtraction and flat fielding, including the rejection of cosmic rays. The spectra were extracted using the apall task, setting two apertures and fixing the centroid of each quasar spectra.

\section{Method} \label{sec:smbh}

As discussed earlier, the SE method combines the BLR line width and size determined from the luminosity to estimate

\begin{equation}
    M_{\text{BH}} = f \frac{R_{\text{BLR}}(\Delta v)^2}{G}
    \label{eq:one}
\end{equation}

\noindent
where R$_{BLR}$ is the distance from the SMBH to the BLR, $\Delta v$ is the virial velocity of the BLR, $G$ is the gravitational constant and $f$ is the virial factor that depends on the unkown kinematics, structure, inclination and distribution of the BLR (\citealt{2004Peterson} and references therein). Since the emission lines may originate under different conditions, the $f$ parameter may differ between them (\citealt{2013shen}), which in turn gives rise to one of the main uncertainties in measuring M$_{\text{BH}}$. The virial factor has been estimated (e.g, \citealt{2006collin,2015woo,2020mediavilla}) from different emission lines. In this paper we assume $f=1$ following the observational constraint given by \citet{2015woo}, which is in agreement with the nonweighted average $\langle f\rangle$ = 0.99 $\pm$ 0.08 given by \citet{2021Mediavilla}. Thanks to the known correlation between the luminosity of the AGN and the size of the BEL (e.g. \citealt{2000kaspi,2005kaspi,2009Bentz}), and assuming viral equilibrium, we estimate the mass as
\begin{eqnarray}
\log( M_{\text{BH}} / M_{\odot}) &= \log(K) \,  + \, \alpha \, \log\left(\frac{\lambda L_{\lambda}}{10^{44} \text{ erg/s }} \right)\nonumber \\
 &  + \, 2.0 \, \log \left(\frac{\text{FWHM}}{1000 \:\text{km/s}} \right),    \label{eq:two}
\end{eqnarray}

\noindent where\\

\noindent
( log $K$ , $\alpha$ ){\textbf{$_{H\alpha}$}} = ( 6.845 , 0.650 ),\\
( log $K$ , $\alpha$ ){\bf $_{H\beta}$} = ( 6.740 , 0.650 ),\\
( log $K$ , $\alpha$ )$_{Mg{\footnotesize{II}}}$ = ( 6.925 , 0.609 ), and\\
( log $K$ , $\alpha$ )$_{C{\footnotesize{IV}}}$ = ( 6.353 , 0.599 )\\

\noindent
are the calibrated parameters from \citet{2018mejia-restrepob} for the H$\alpha$, H$\beta$, Mg{\footnotesize{II}} and C{\footnotesize{IV}} lines, respectively, and the luminosities are those at 5100\AA~(L$_{5100}$) for H$\alpha$ and H$\beta$, 3000\AA~(L$_{3000}$) for Mg{\footnotesize{II}}, and 1450\AA~(L$_{1450}$) for C{\footnotesize{IV}}.

\subsection{Emission line fitting}

We modeled the emission line profiles after removing the continuum and an iron line template, following \citet{2016Mejiarestrepo}. We use a maximum of two Gaussian broad components and a single narrow line component for each emission line. In addition to the narrow and broad components of the principal emission lines (H$\alpha$, H$\beta$, C{\footnotesize{IV}} and Mg{\footnotesize{II}}), we added four extra components in the H$\alpha$ profile for the [N II] and [S II] narrow-line doublets, two for the [O III] NLR doublet in the H$\beta$ profile plus one to the He II broad emission line. We masked regions with telluric absorption problems, bad seeing and poor S/N that could affect our fit. The best final fit is shown as a red line in Figure~\ref{fig:mbh} for the LUCIFER and MMIRS data, and in \ref{FigmbhQJ0158,FigmbhLBQS1333,FigmbhQ1355,Figmbh1226}
for QJ0158$-$4325, LBQS1333$-$0113, Q1355$-$2257 and SDSS1226$-$0006, respectively. In the cases using two broad emission lines, the FWHM was calculated from the combined profile after removing the NLR components. We carried out a Monte Carlo simulation consisting of 1000 simulated spectra randomnly adding the estimated spectral noise to obtain a 95$\%$ confidence uncertainty estimate.

\begin{figure*}[!htbp]
   \centering
   \vspace{-1\baselineskip}
    \begin{subfigure}[b]{1.0\textwidth}
 \includegraphics[width=0.44\textwidth]{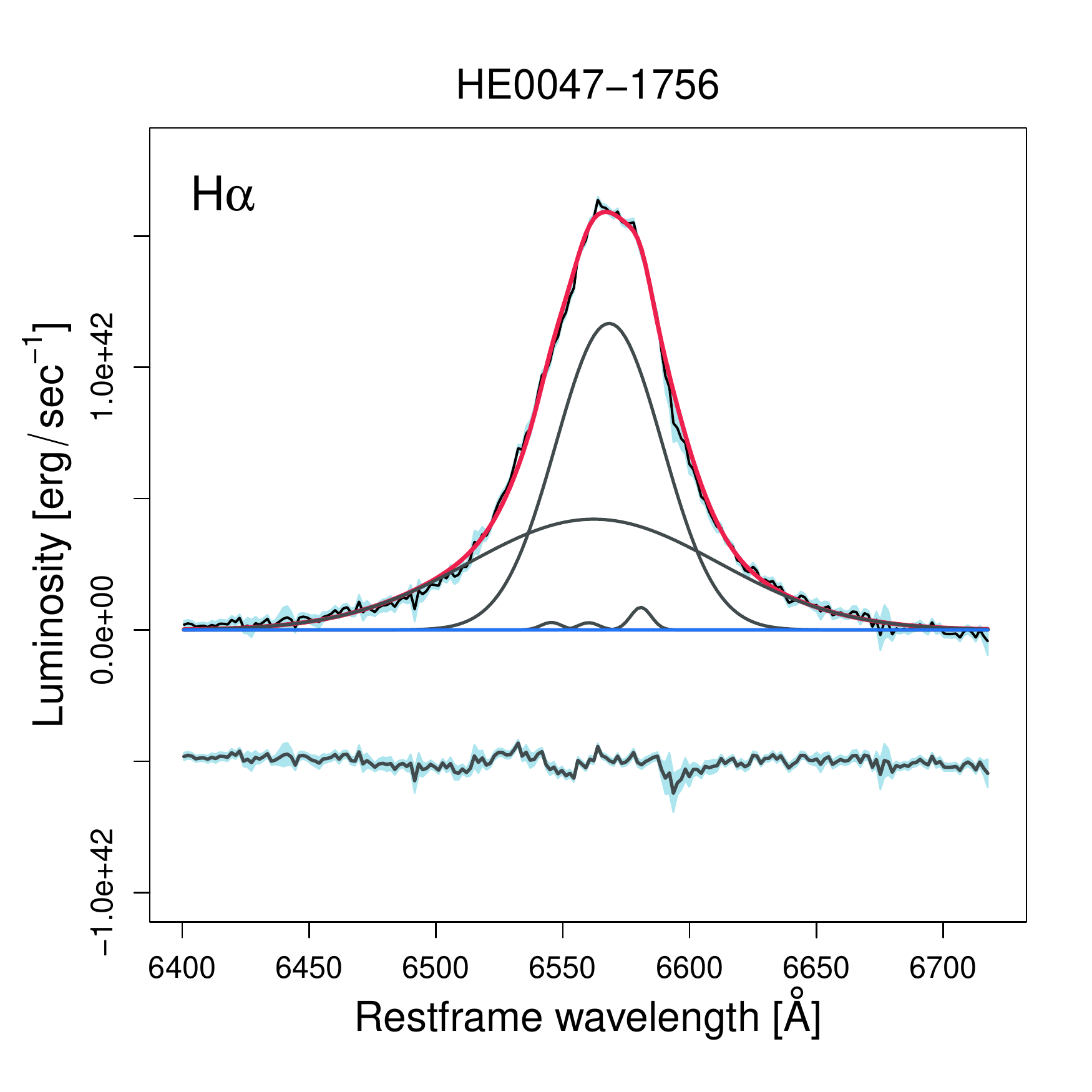}
    \includegraphics[width=0.44\textwidth]{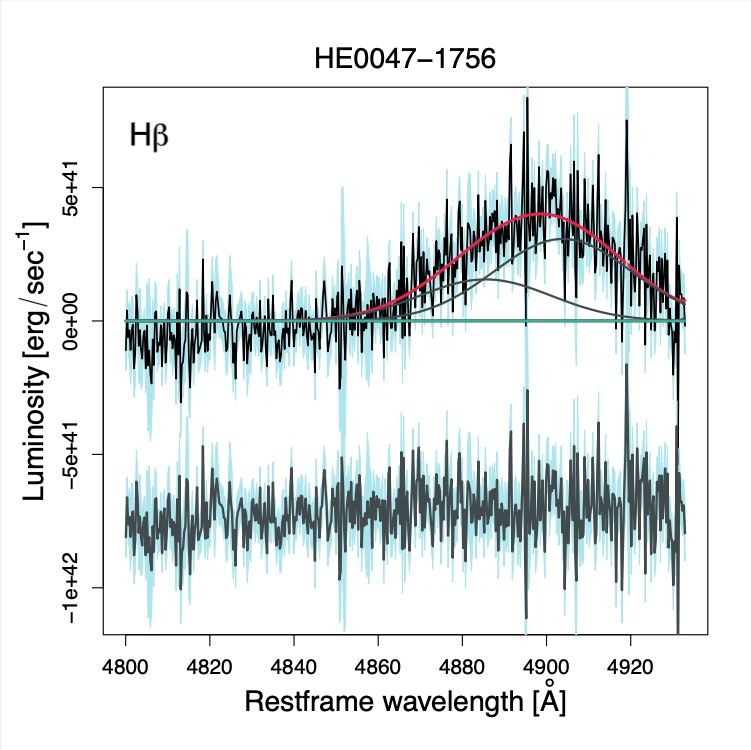}
    \end{subfigure}
    \vspace{-1\baselineskip}
    \begin{subfigure}{1.0\textwidth}
  \includegraphics[width=0.44\textwidth]{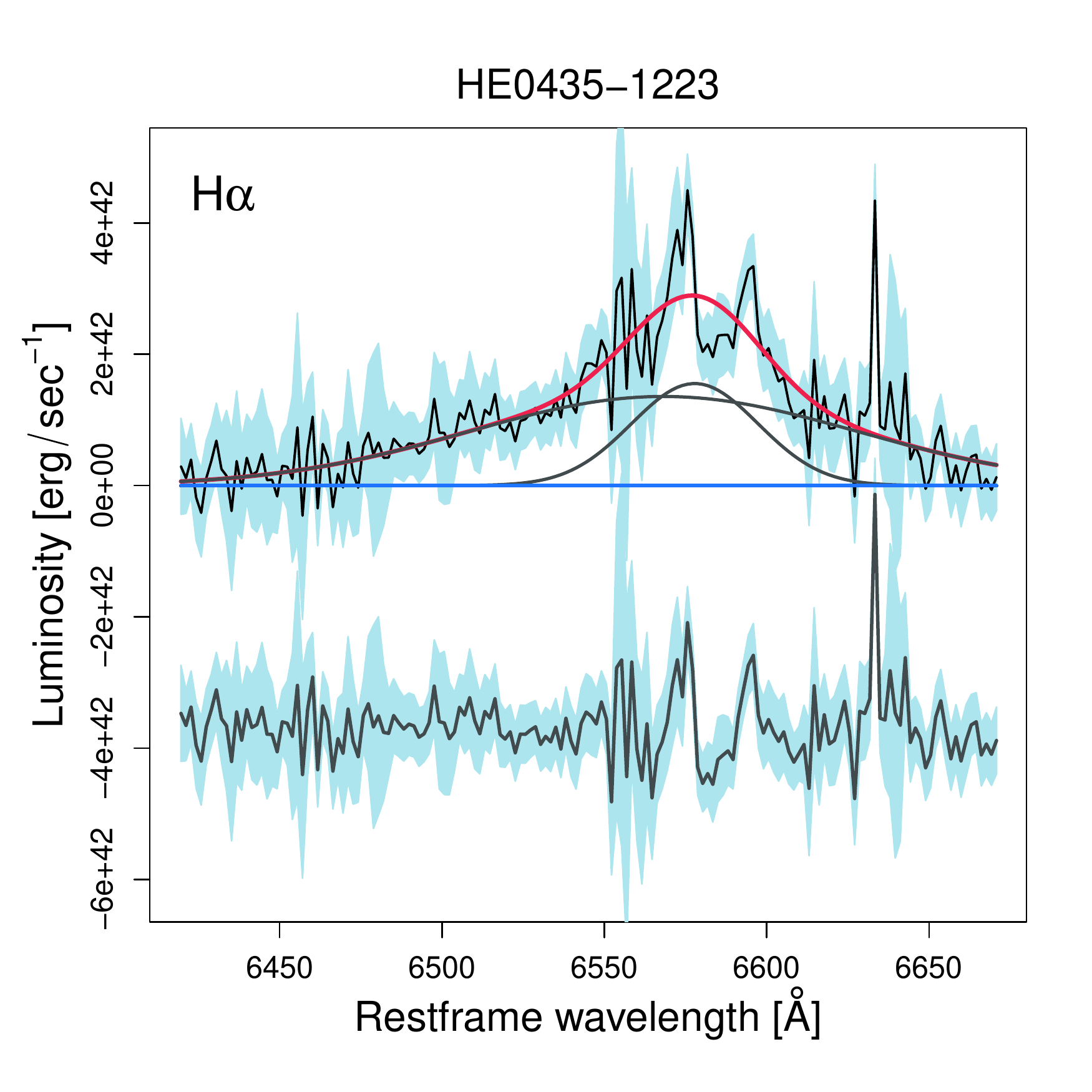}
  \includegraphics[width=0.44\textwidth]{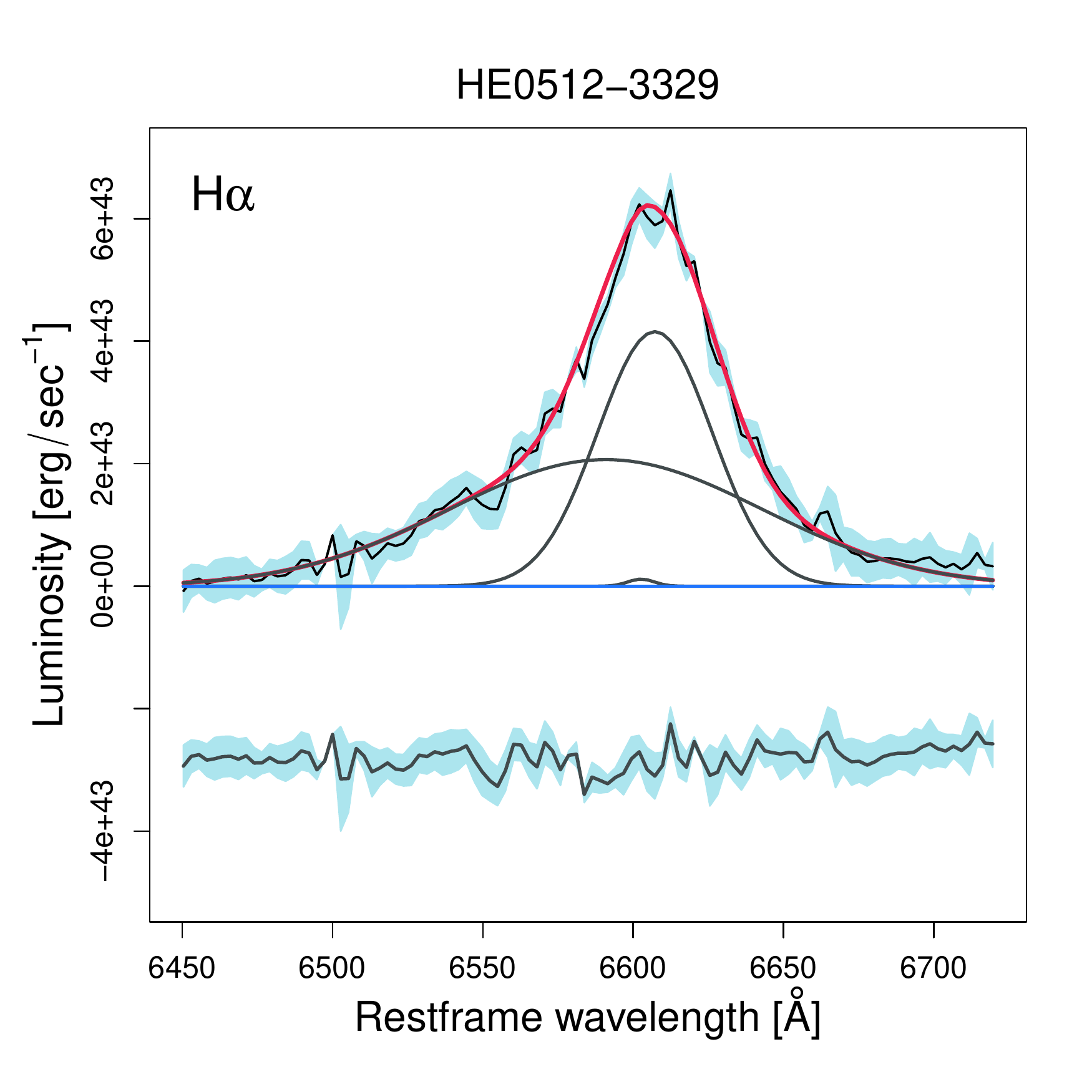}
    \end{subfigure}
    \begin{subfigure}{1.0\textwidth}
  \includegraphics[width=0.44\textwidth]{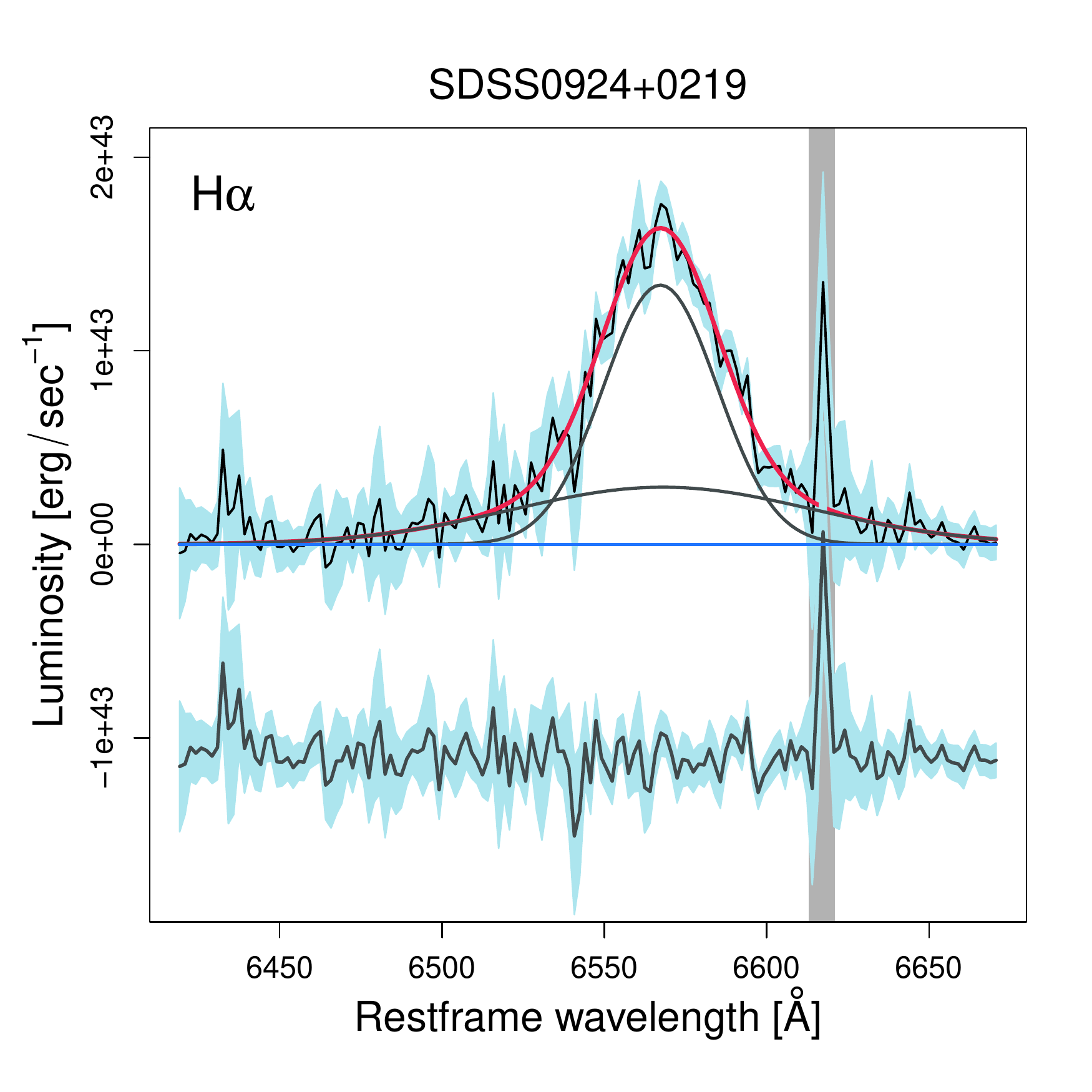}
  \includegraphics[width=0.44\textwidth]{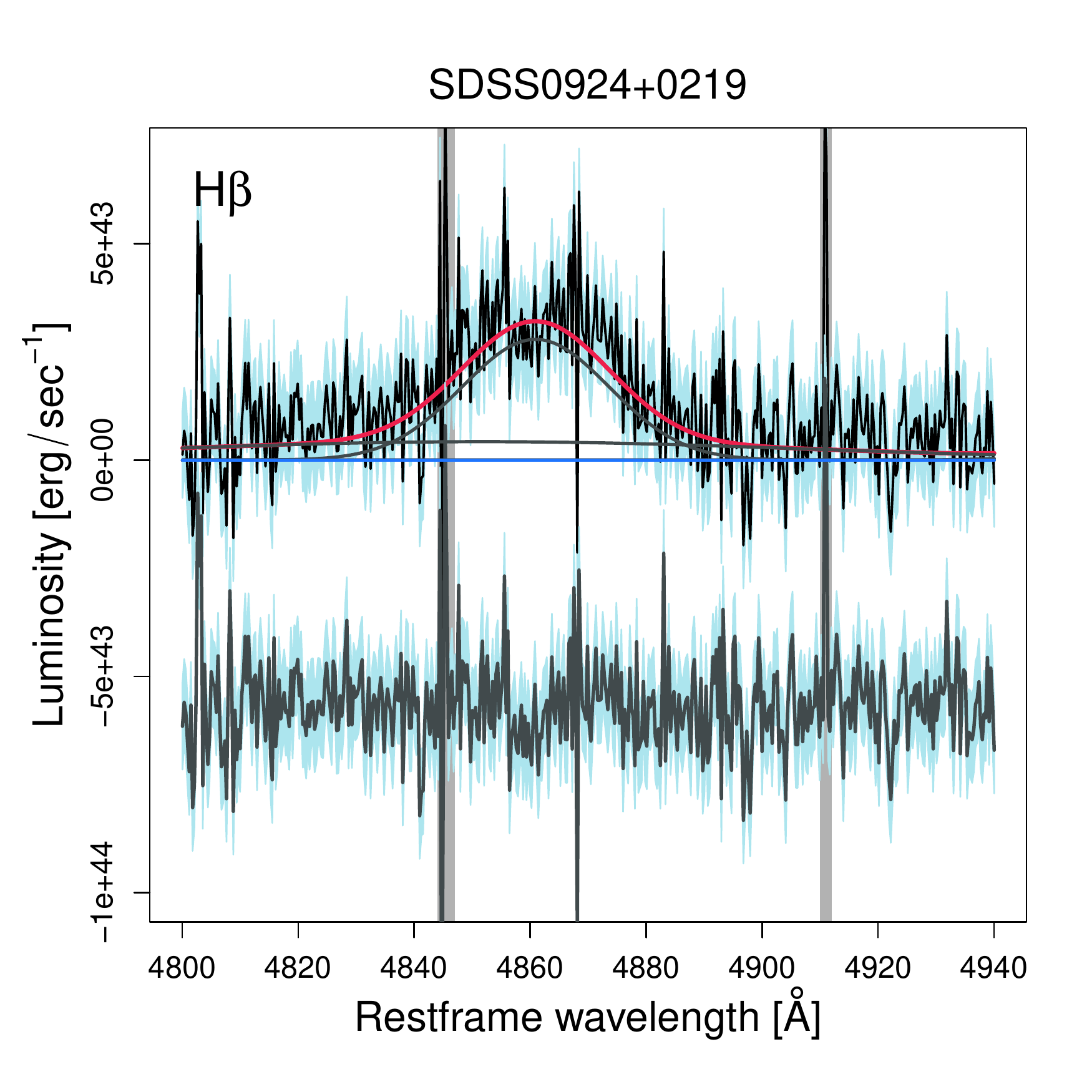}
    \end{subfigure}
      \caption{Gaussian fits to the H$\alpha$ and H$\beta$ lines of the lensed systems. The red line is the best fit, the black lines are the different components of each region (emission and absorption), the green line is the Fe template and the blue line is the continuum fit. The 1-sigma errors are shown by the blue regions and the model residuals are shown below each spectrum.}
\end{figure*}

\begin{figure*}[!htbp]
\ContinuedFloat
\vspace{-1\baselineskip}
    \begin{subfigure}{1.0\textwidth}
  \includegraphics[width=0.44\textwidth]{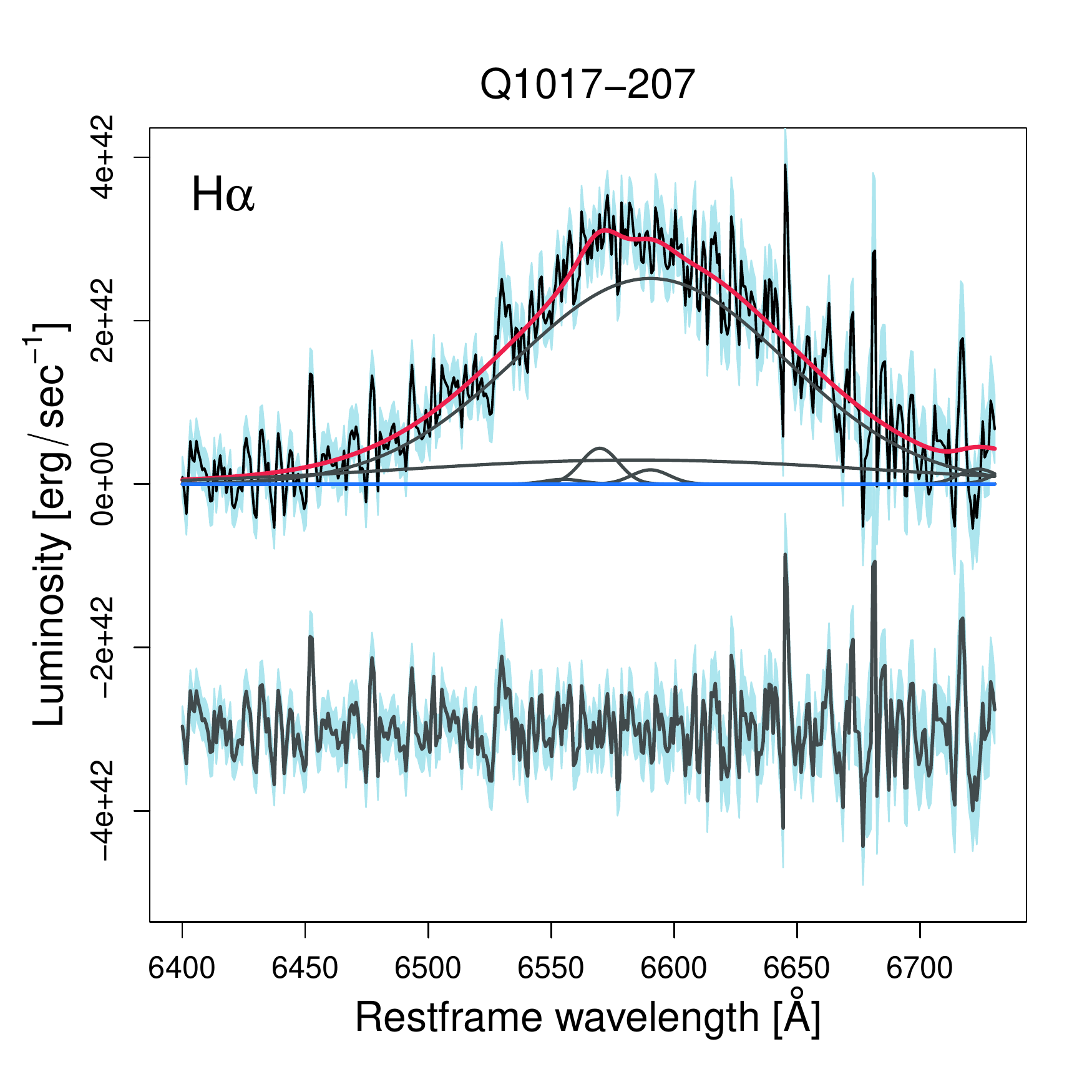}
   \includegraphics[width=0.44\textwidth]{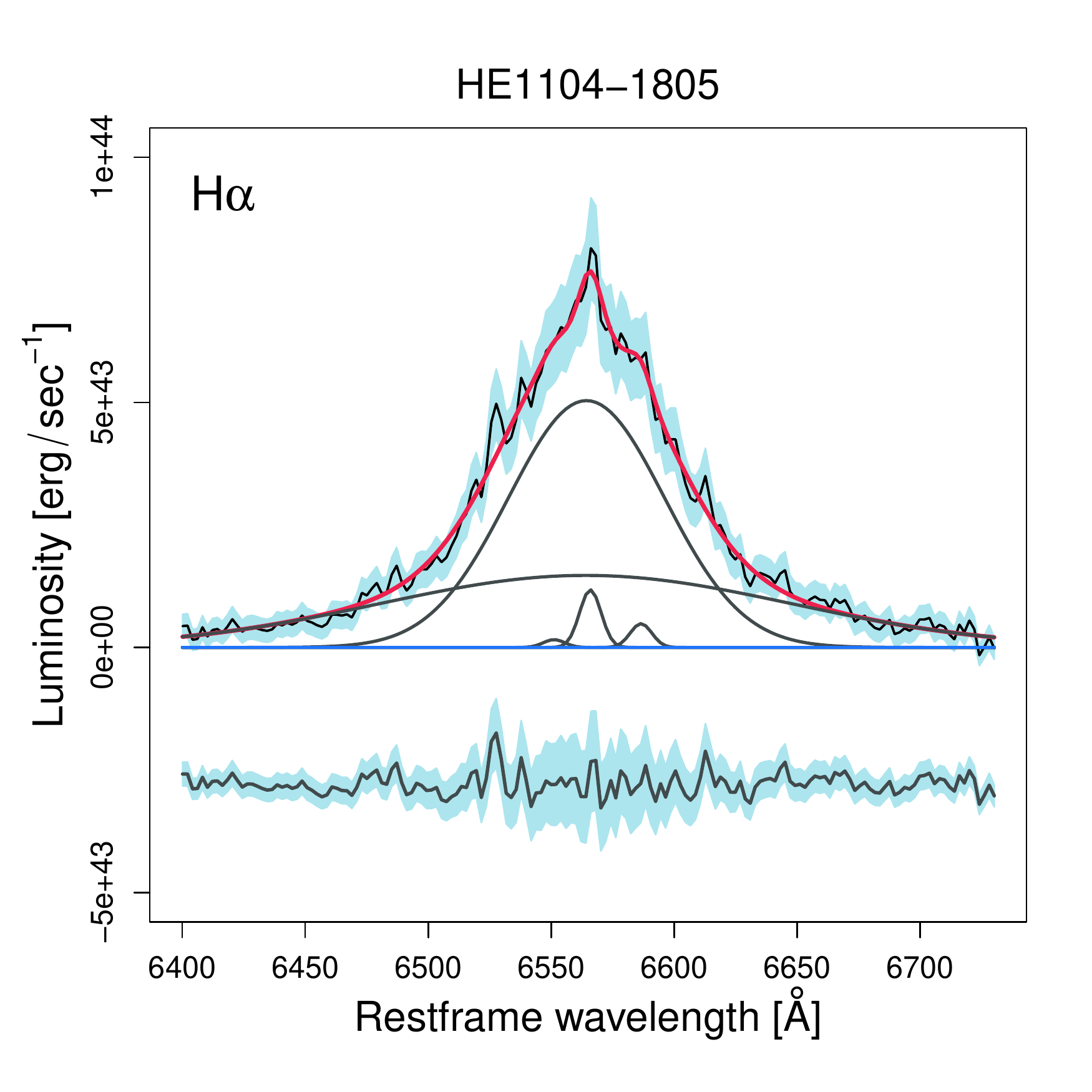}
    \end{subfigure}
\vspace{-1\baselineskip}
    \begin{subfigure}{1.0\textwidth}
  \includegraphics[width=0.44\textwidth]{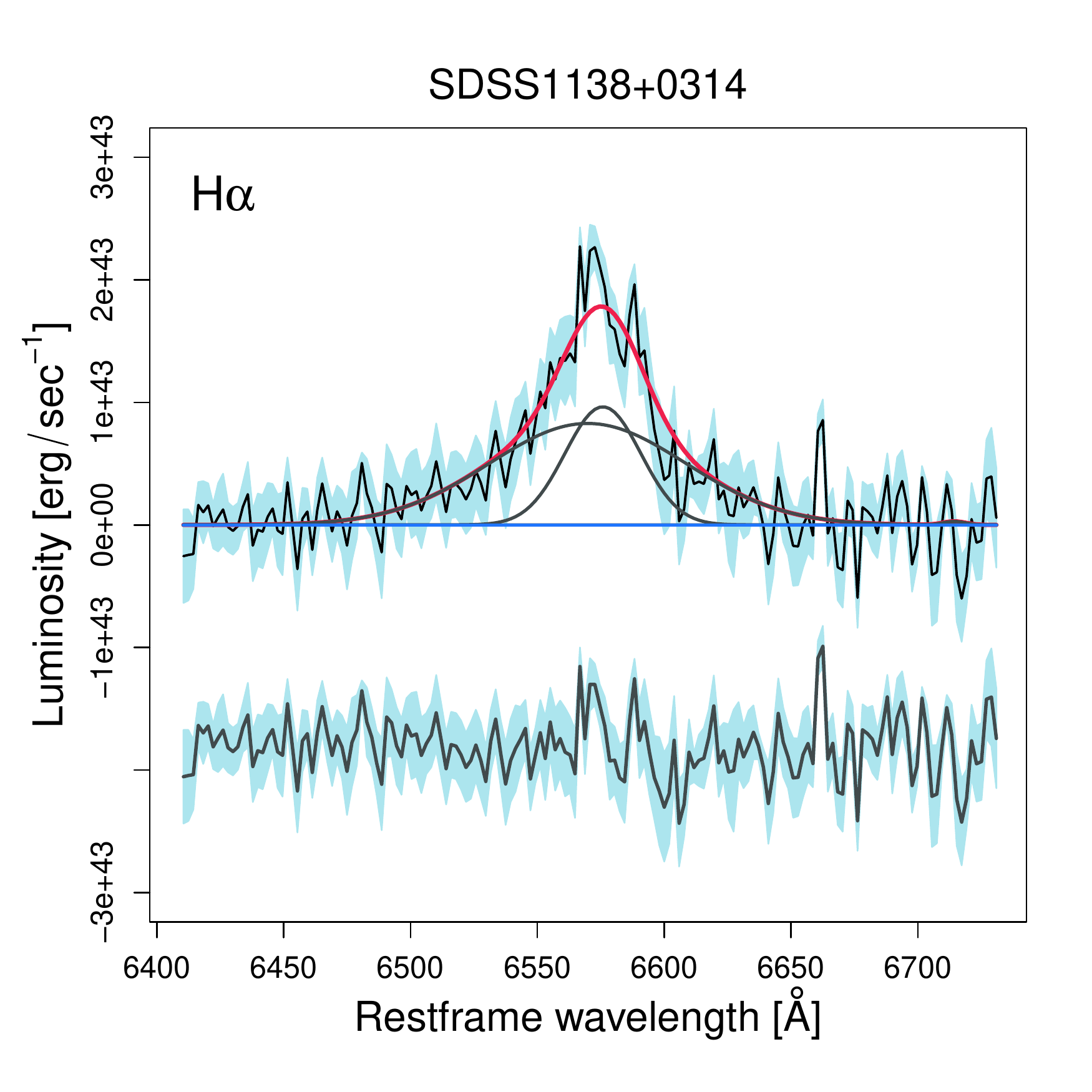}
  \includegraphics[width=0.44\textwidth]{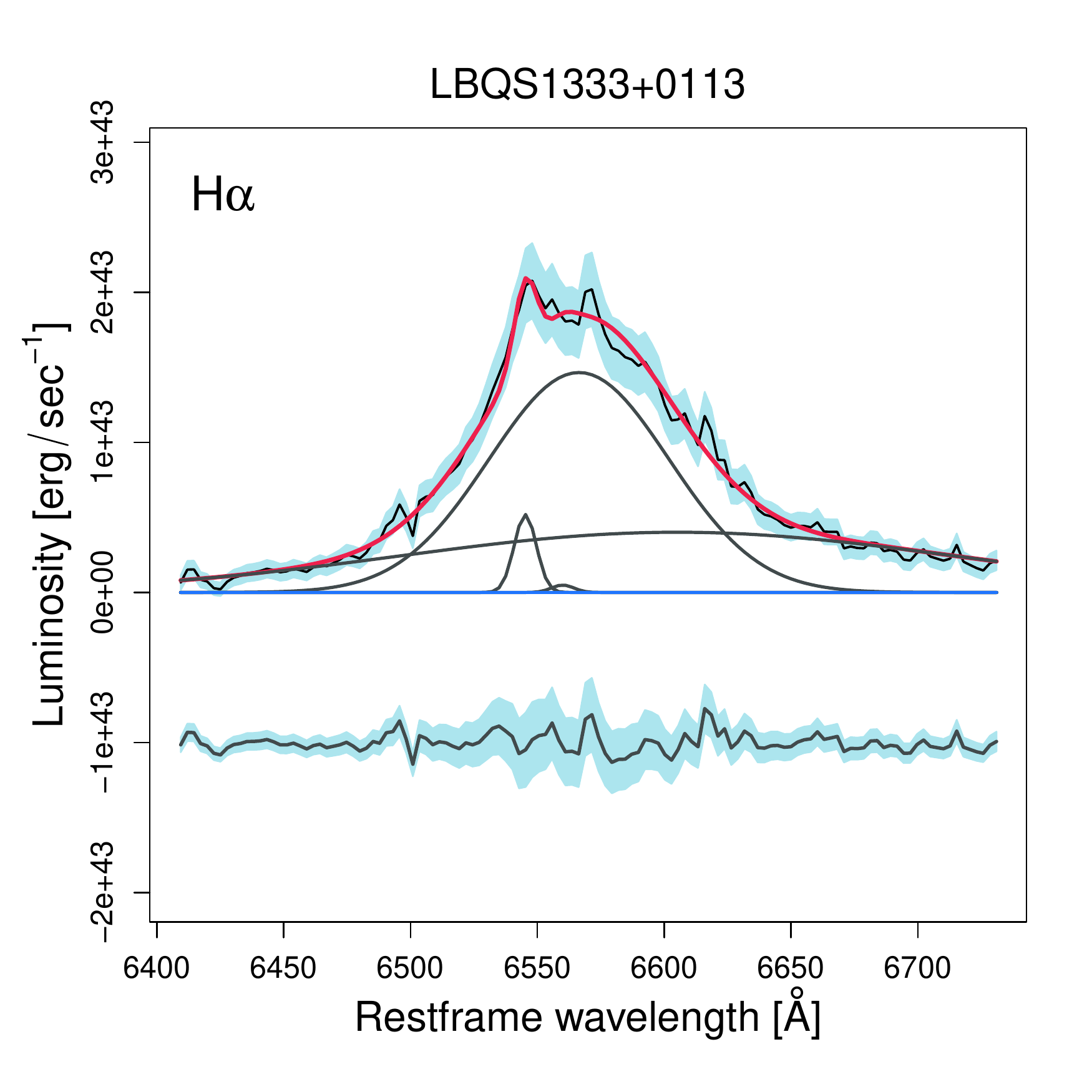}
    \end{subfigure}
    \hfill
    \begin{subfigure}{1.0\textwidth}
  \includegraphics[width=0.44\textwidth]{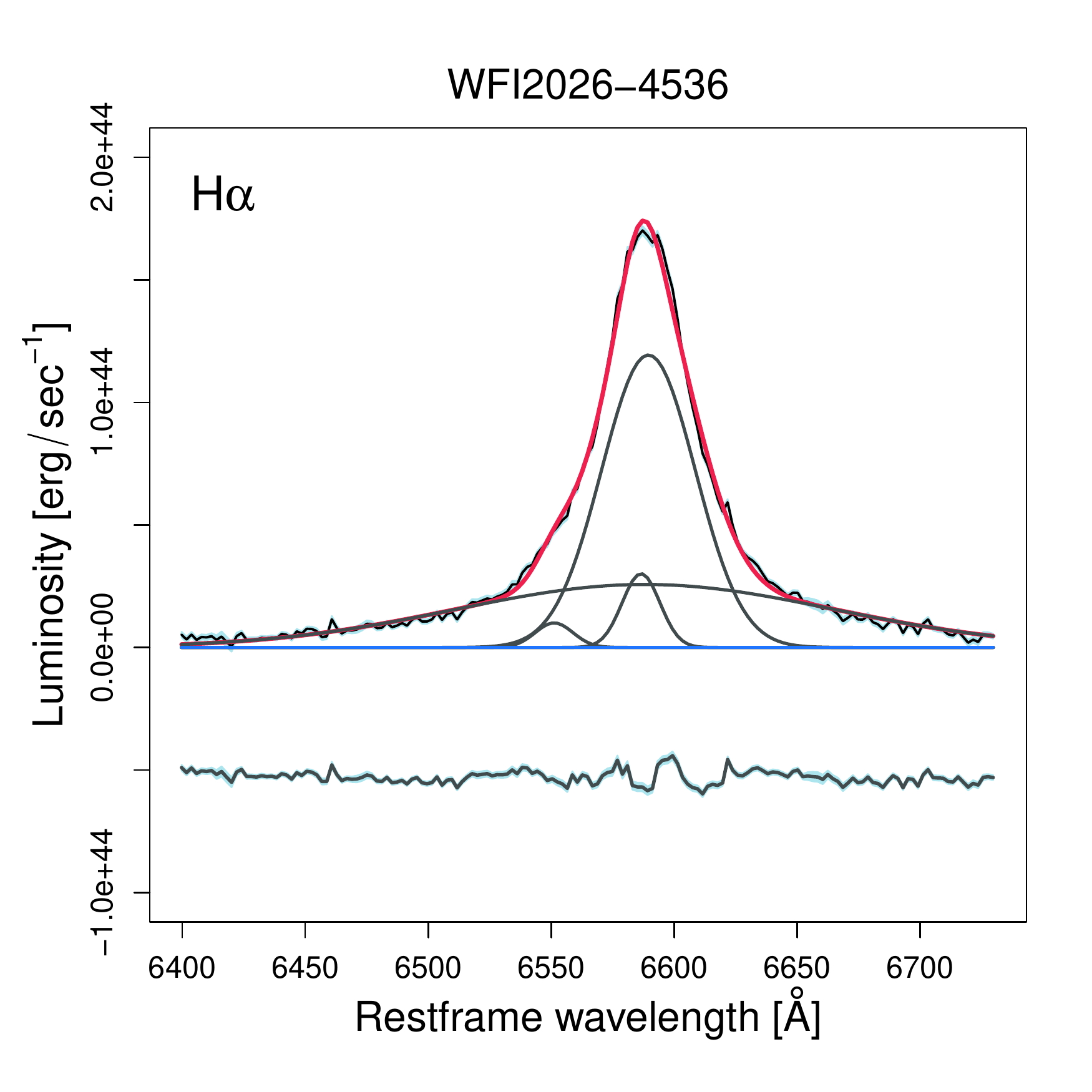}
  \includegraphics[width=0.44\textwidth]{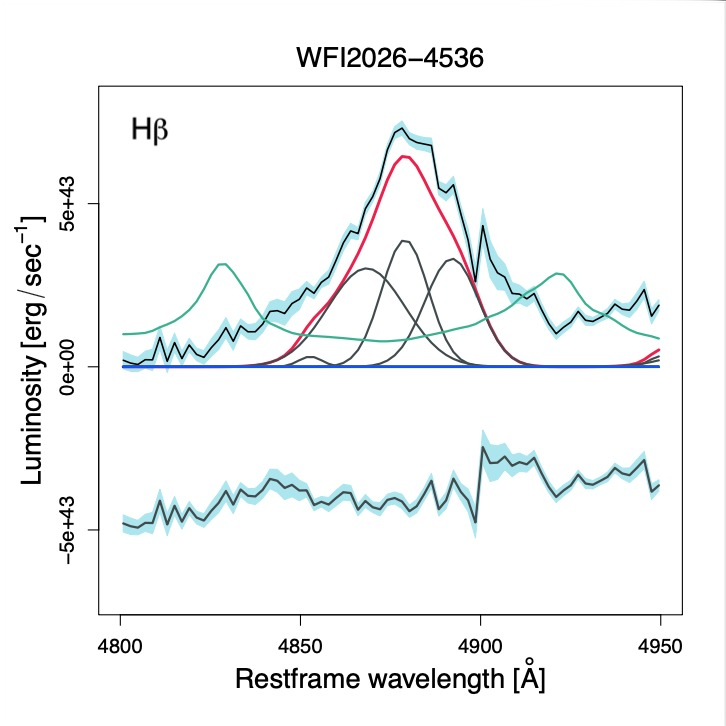}
    \end{subfigure}
      \caption{(cont) Gaussian fits to the H$\alpha$ and H$\beta$ lines of the lensed systems. The red line is the best fit, the black lines are the different components of each region (emission and absorption), the green line is the Fe template and the blue line is the continuum fit. The 1-sigma errors are shown by the blue regions and the model residuals are shown below each spectrum.}
  \label{fig:mbh}
\end{figure*}

\begin{figure*}[!htbp]
\ContinuedFloat
    \begin{subfigure}{1.0\textwidth}
  \includegraphics[width=0.44\textwidth]{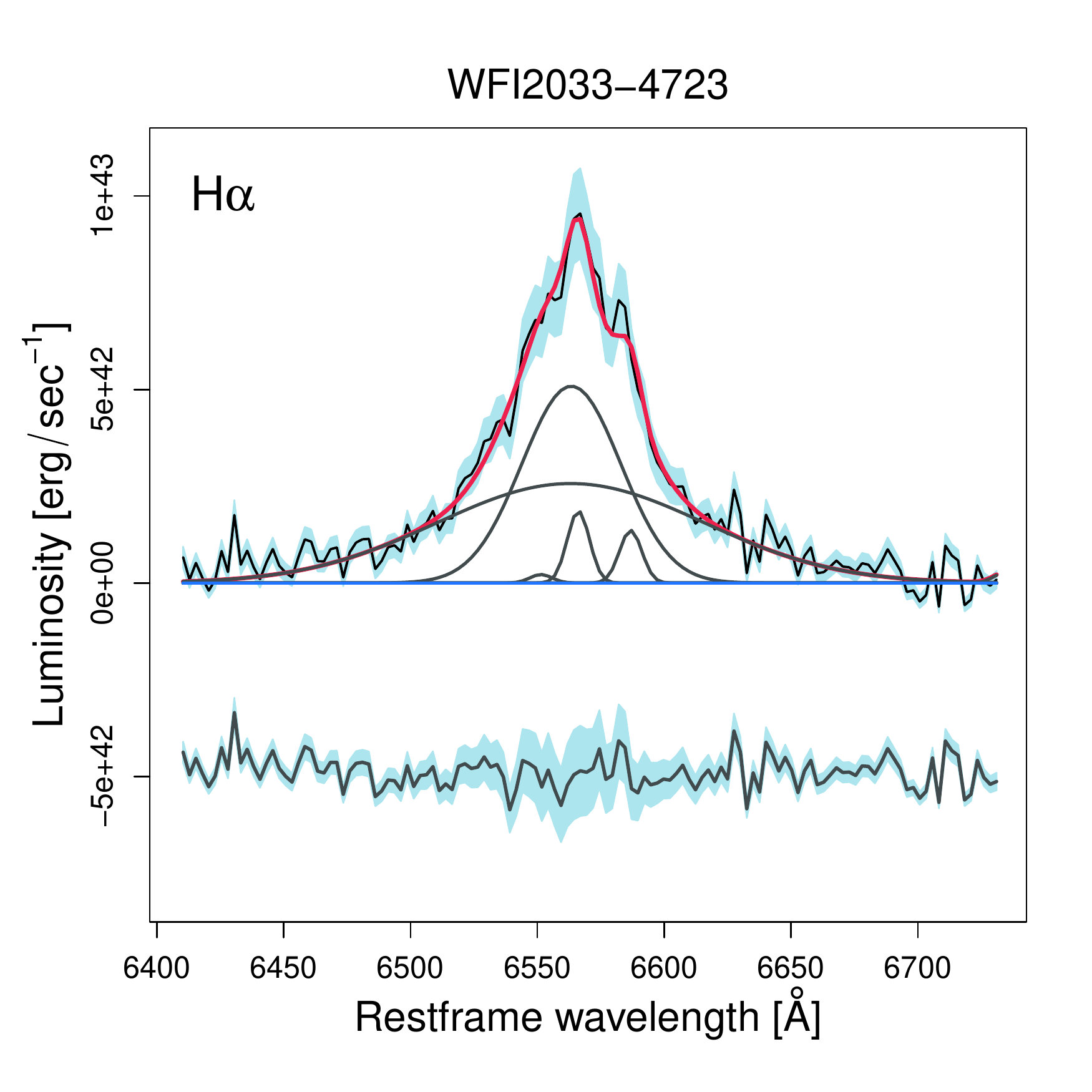}
   \includegraphics[width=0.44\textwidth]{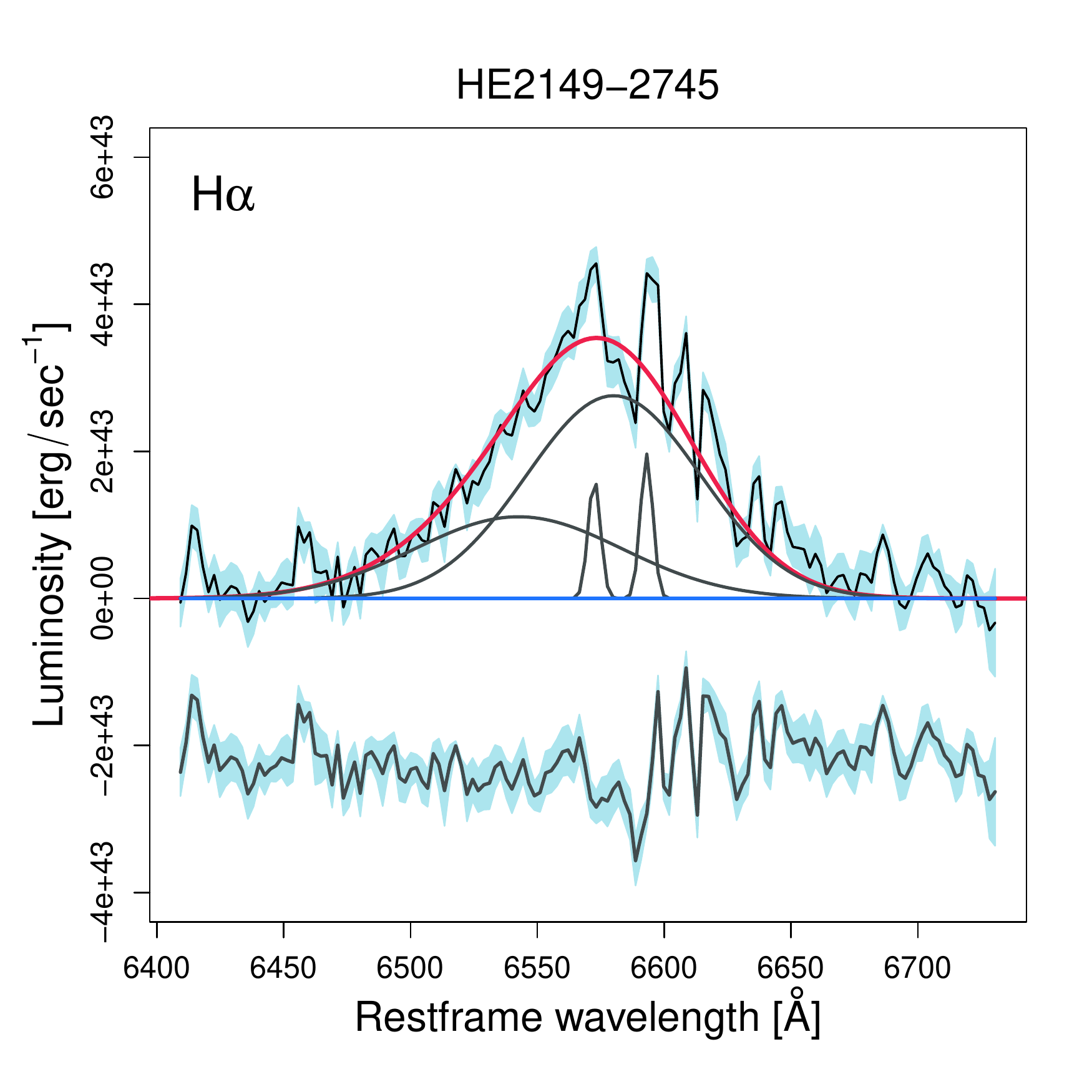}
    \end{subfigure}
      \caption{(cont) Gaussian fits to the H$\alpha$ and H$\beta$ lines of the lensed systems. The red line is the best fit, the black lines are the different components of each region (emission and absorption), the green line is the Fe template and the blue line is the continuum fit. The 1-sigma errors are shown by the blue regions and the model residuals are shown below each spectrum.}
  \label{fig:mbh}
\end{figure*}


\begin{figure*}[!htbp]
    \begin{subfigure}{1.0\textwidth}
  \includegraphics[width=0.43\textwidth]{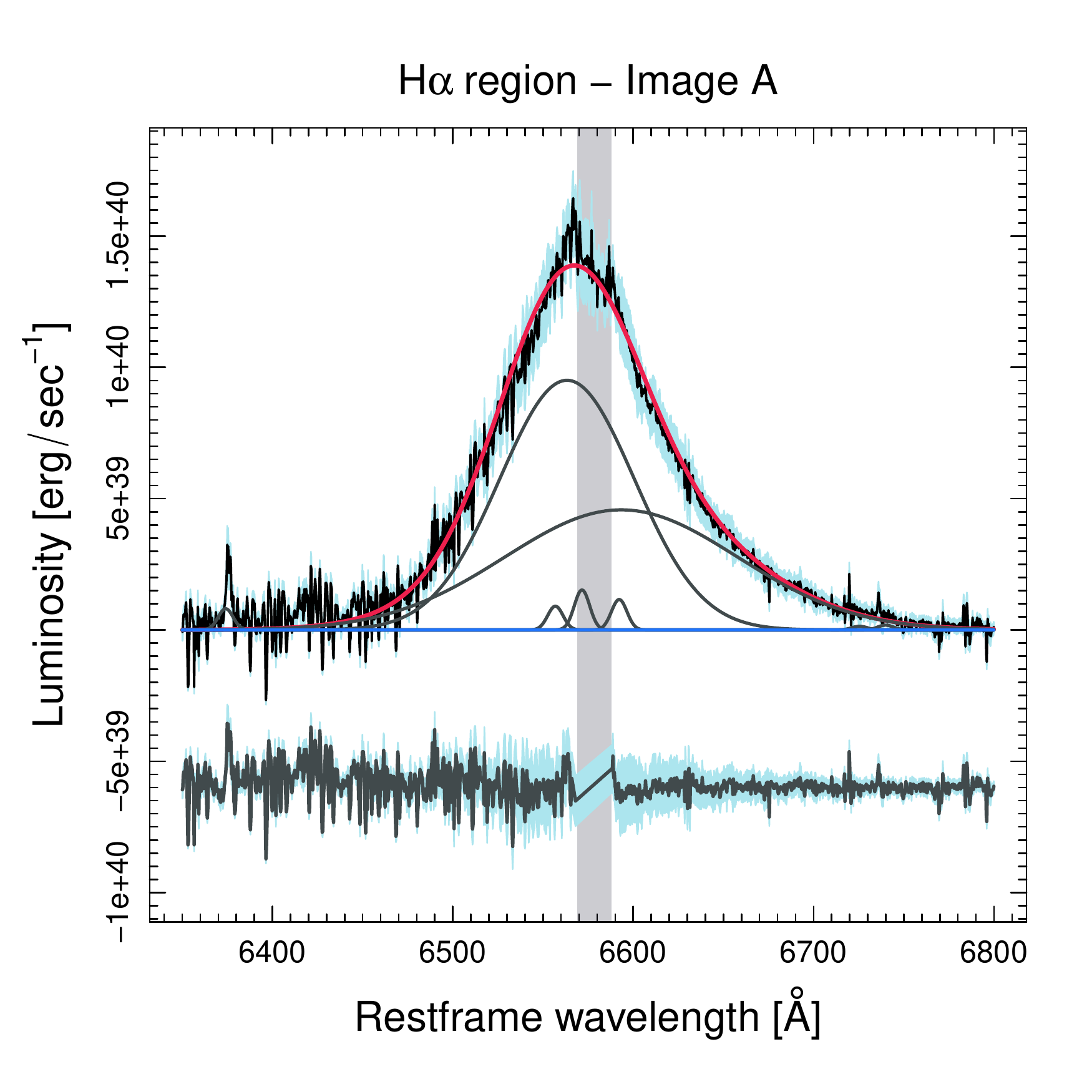}
   \includegraphics[width=0.43\textwidth]{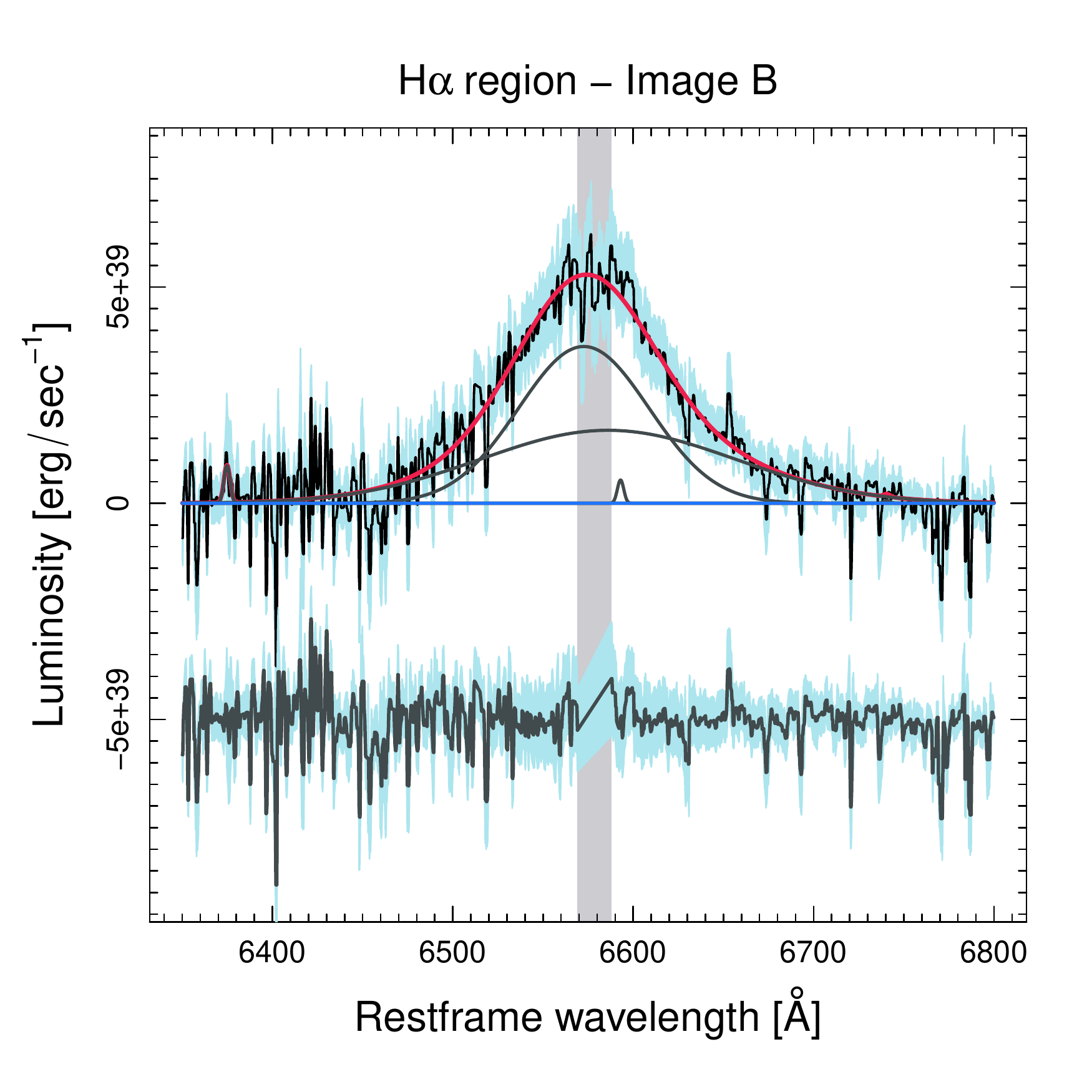}
    \end{subfigure}
        \hfill
    \begin{subfigure}{1.0\textwidth}
  \includegraphics[width=0.43\textwidth]{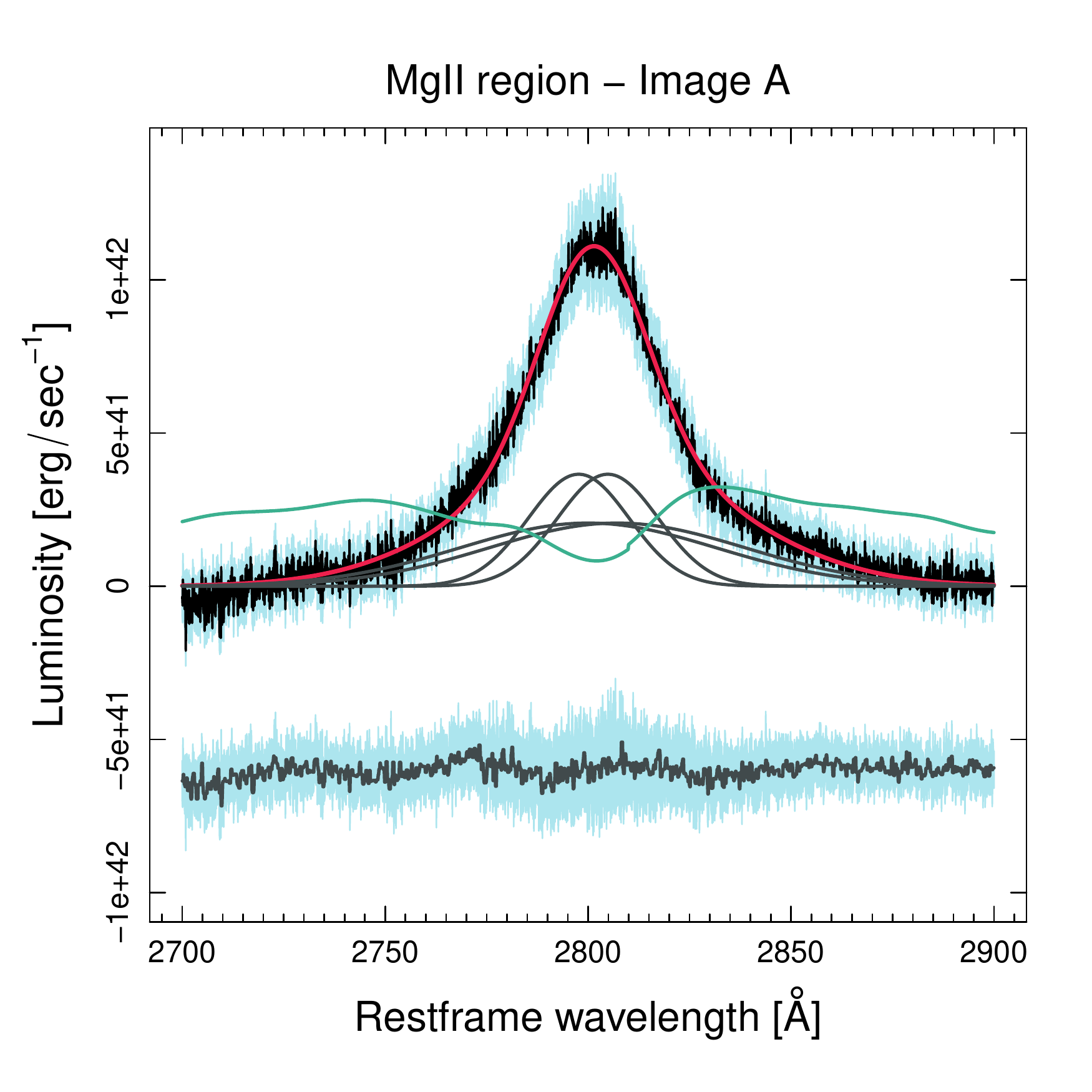}
  \includegraphics[width=0.43\textwidth]{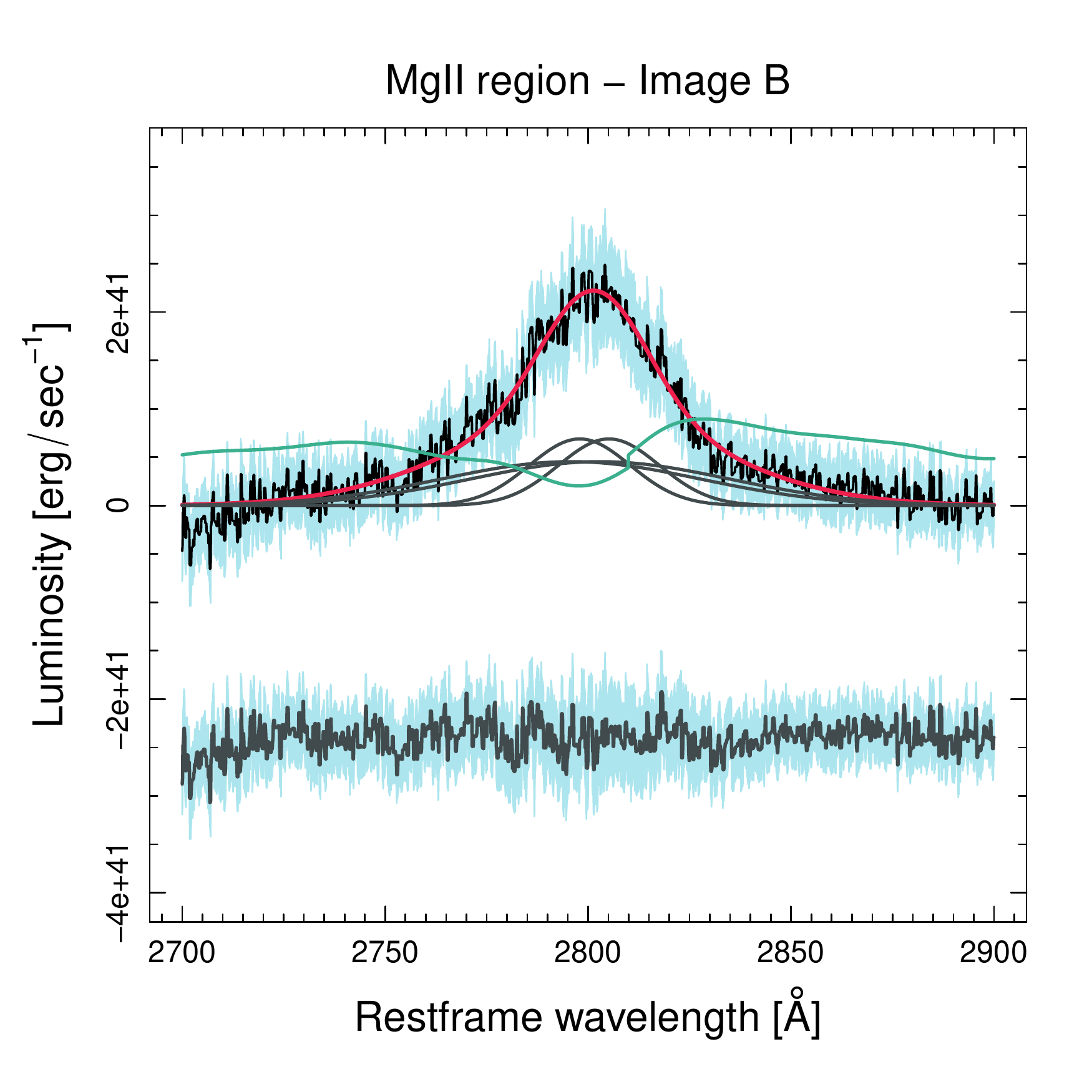}
    \end{subfigure}
    \hfill
    \begin{subfigure}{1.0\textwidth}
  \includegraphics[width=0.43\textwidth]{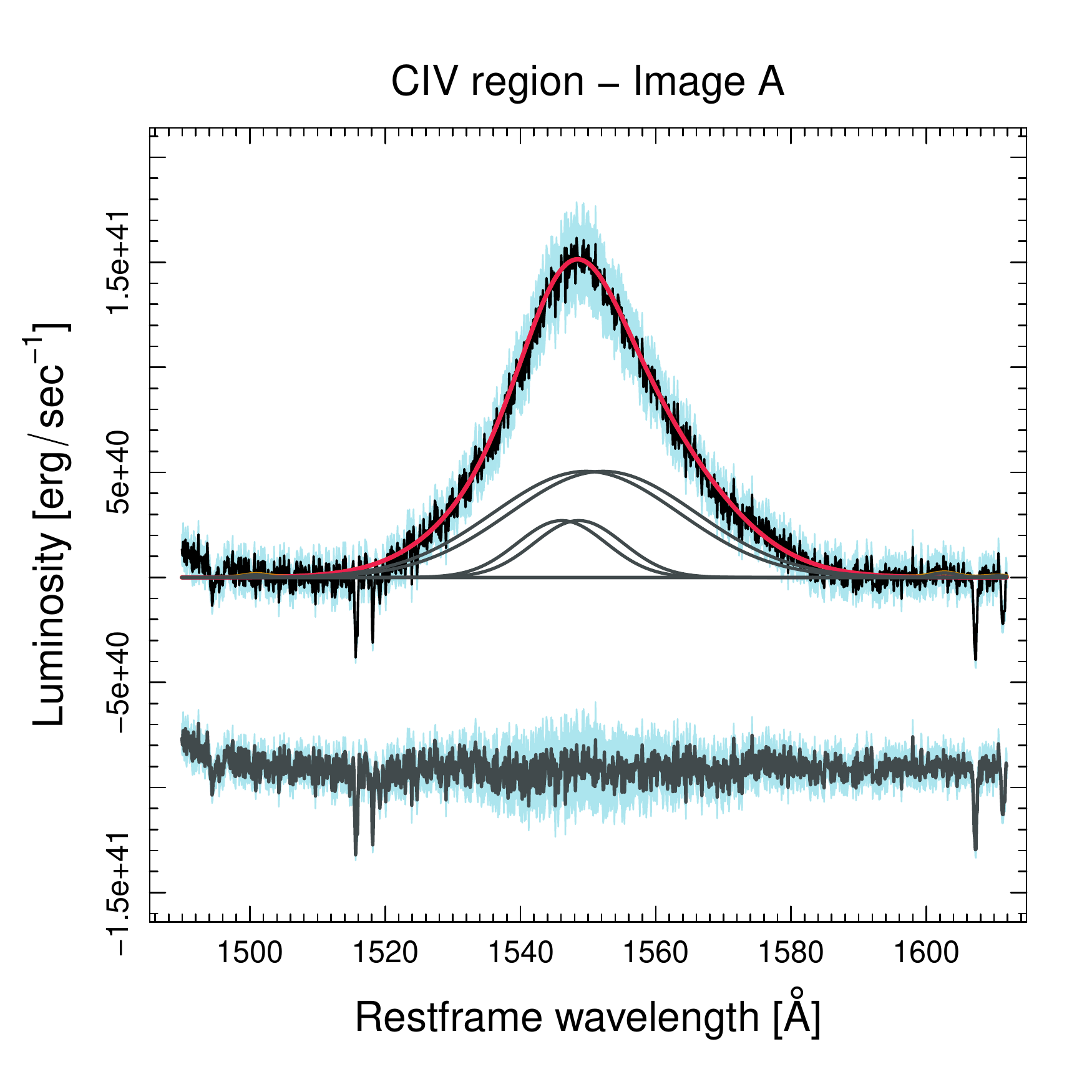}
  \includegraphics[width=0.43\textwidth]{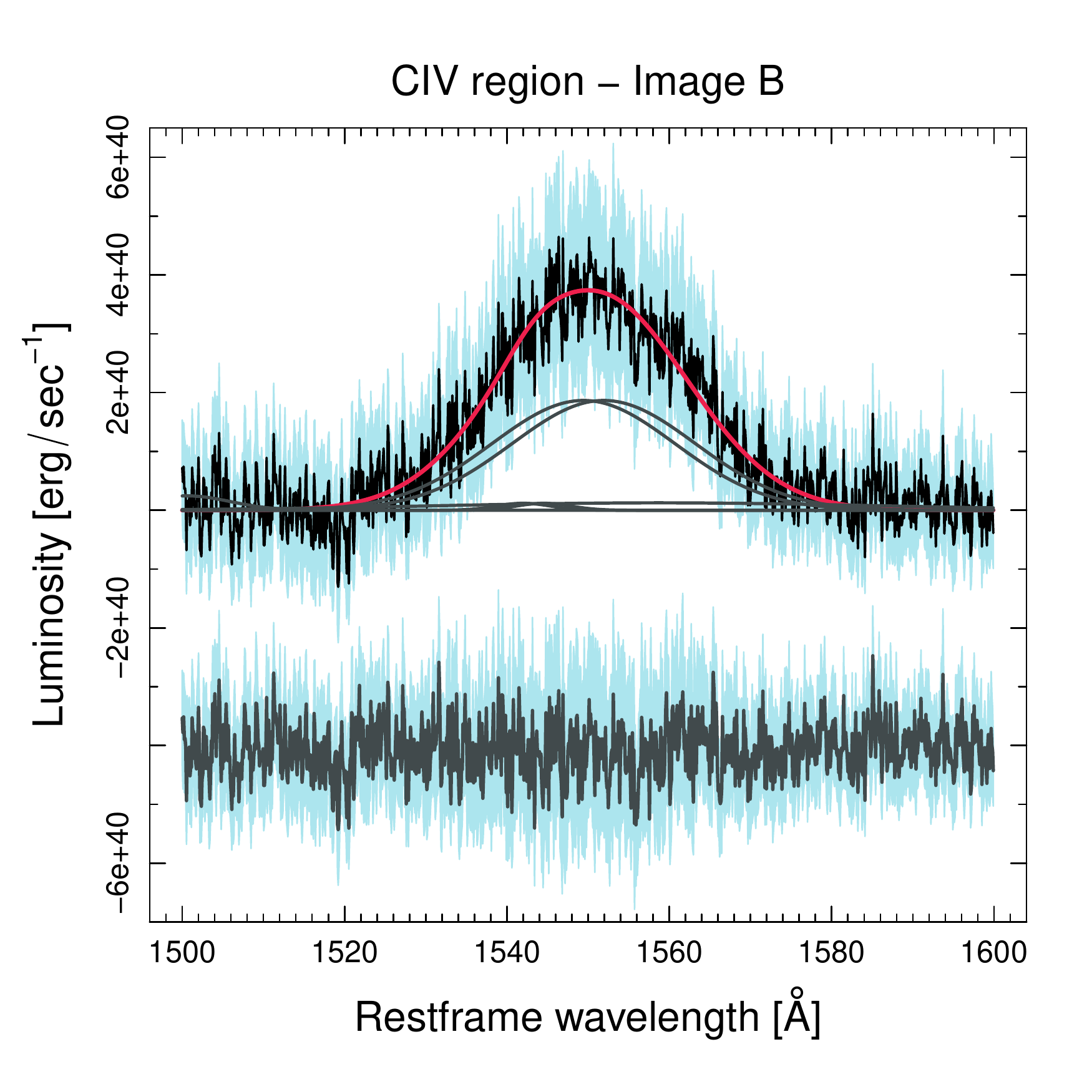}
    \end{subfigure}
      \caption{Gaussian fits to the A and B image broad emission lines of QJ0158-4325. The red line is the best fit, the black lines are the different components of each region (emission and absorption), the green line is the Fe template and the blue line is the continuum fit. The 1-sigma errors are shown by the blue regions and the model residuals are shown below each spectrum.}
  \label{FigmbhQJ0158}
\end{figure*}


\begin{figure*}[!htbp]
    \begin{subfigure}{1.0\textwidth}
  \includegraphics[width=0.44\textwidth]{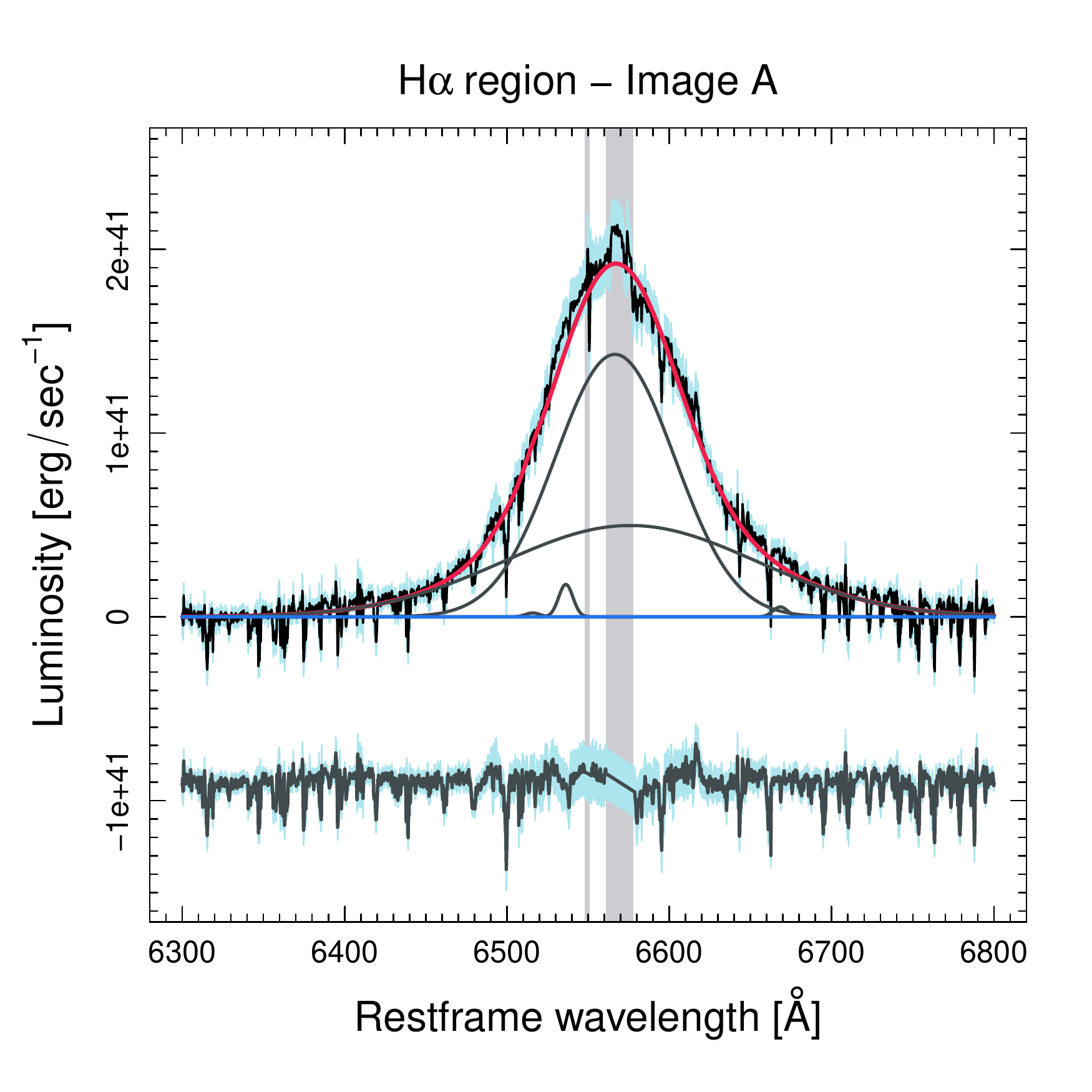}
   \includegraphics[width=0.44\textwidth]{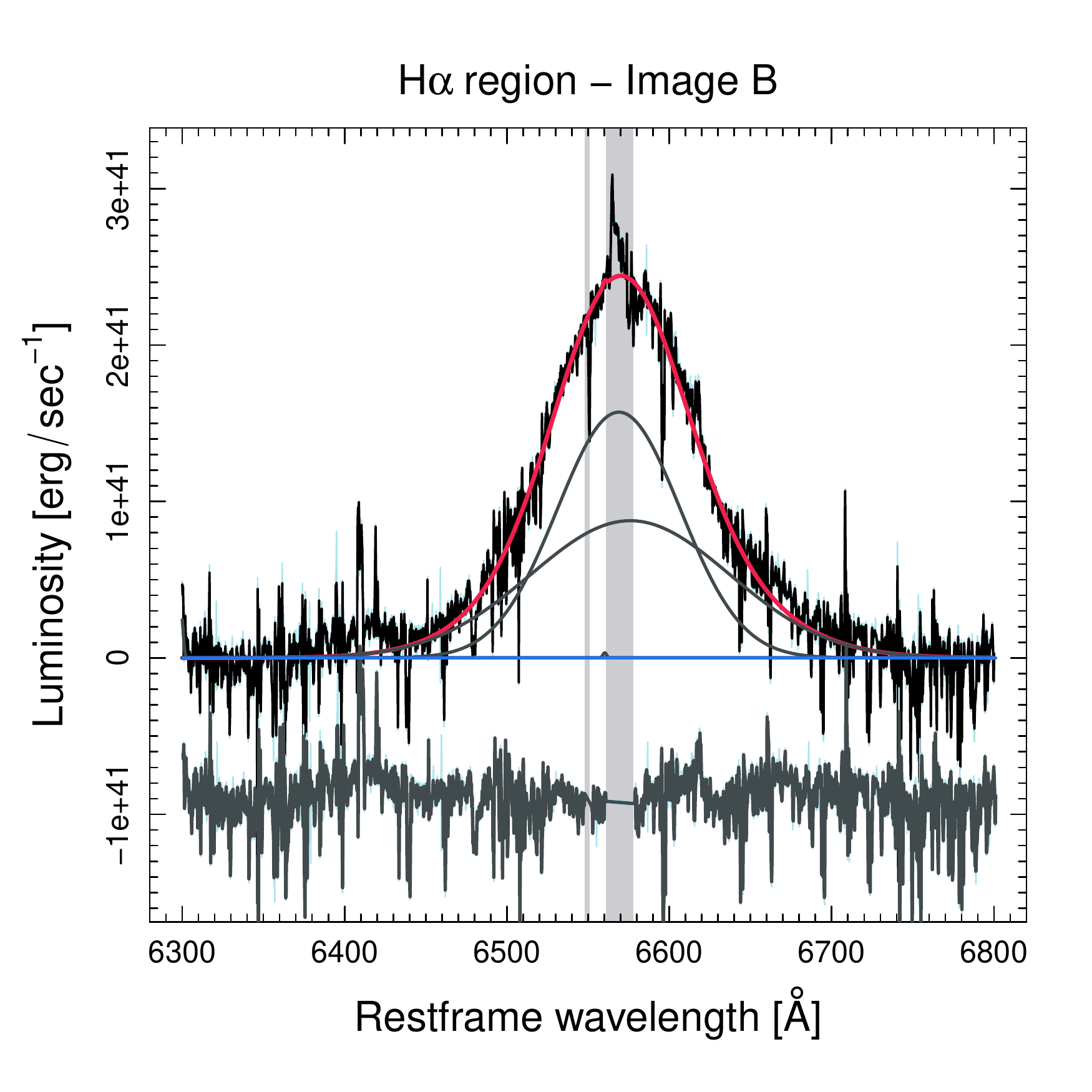}
    \end{subfigure}
        \hfill
    \begin{subfigure}{1.0\textwidth}
  \includegraphics[width=0.44\textwidth]{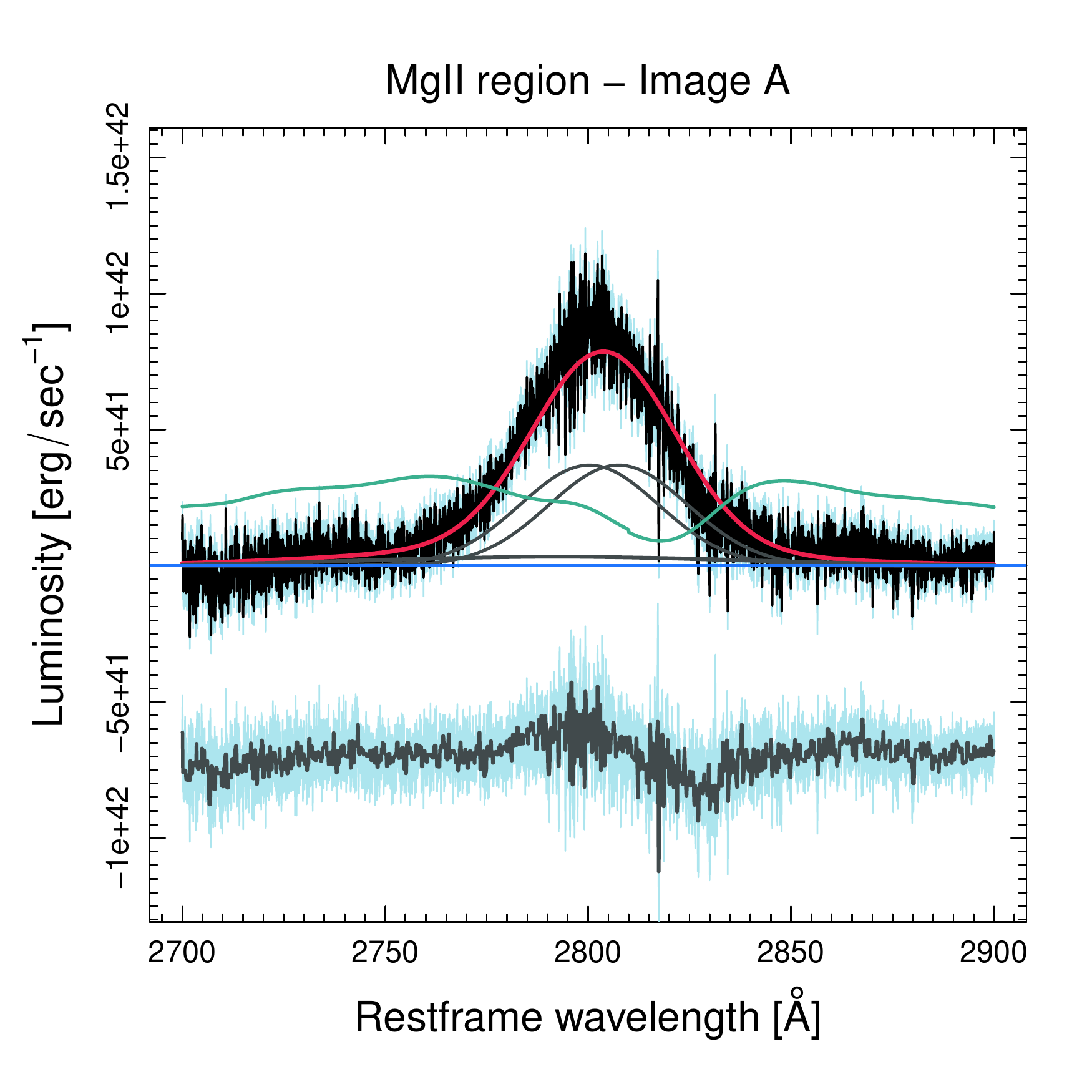}
  \includegraphics[width=0.44\textwidth]{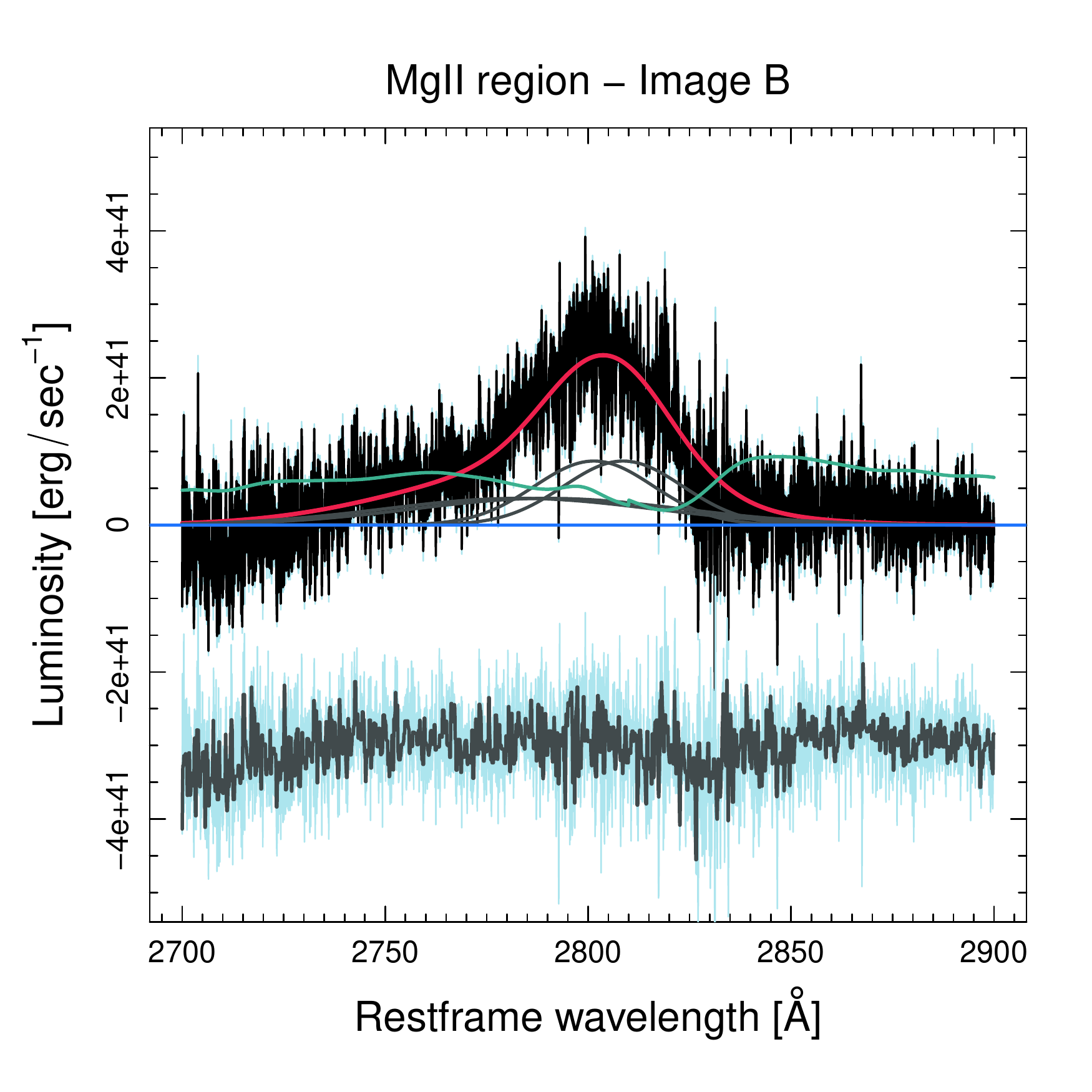}
    \end{subfigure}
      \caption{Gaussian fits to the A and B image broad emission lines of LBQS1333+0113. The red line is the best fit, the black lines are the different components of each region (emission and absorption), the green line is the Fe template and the blue line is the continuum fit. The 1-sigma errors are shown by the blue regions and the model residuals are shown below each spectrum.}
  \label{FigmbhLBQS1333}
\end{figure*}


\begin{figure*}[!htbp]
    \begin{subfigure}{1.0\textwidth}
  \includegraphics[width=0.43\textwidth]{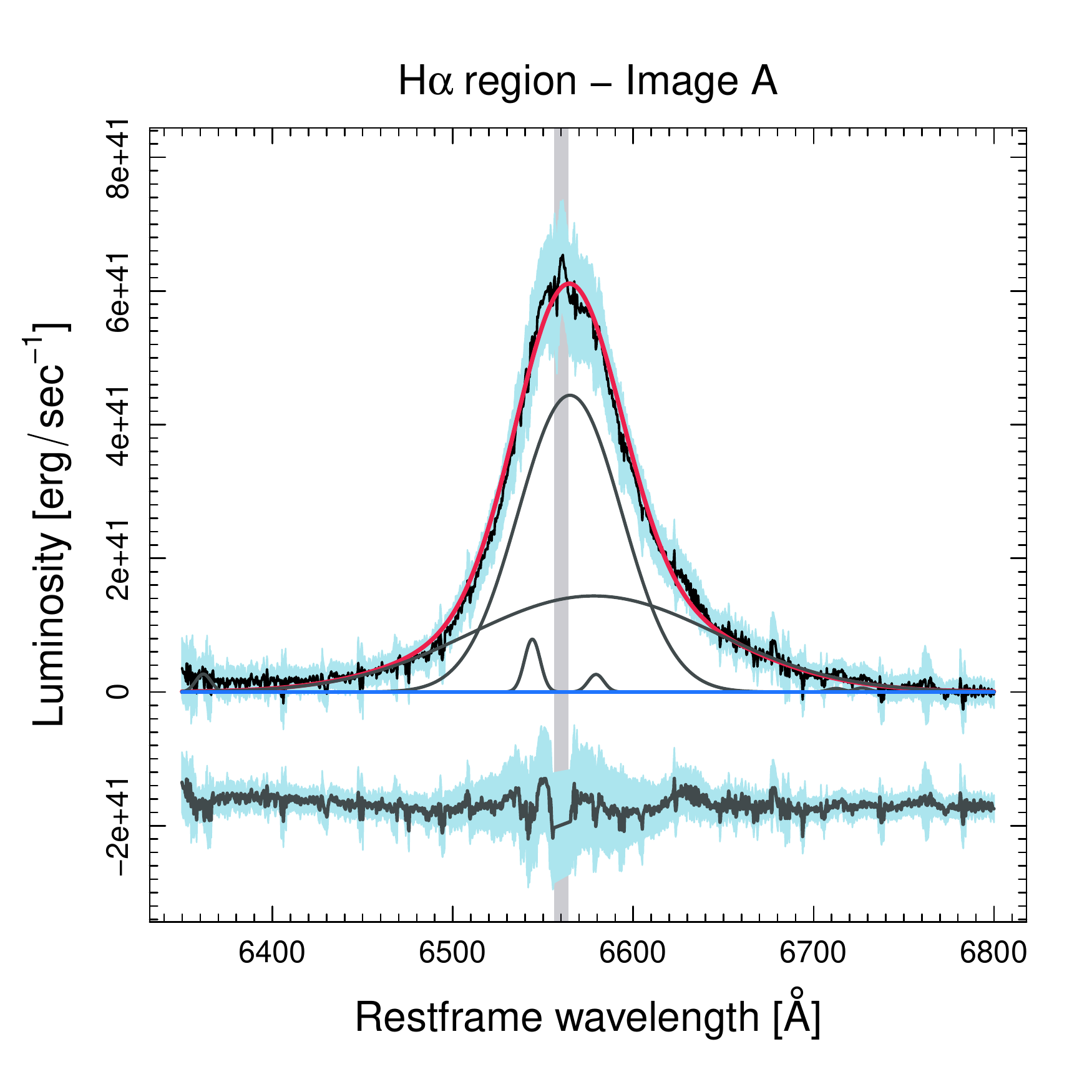}
   \includegraphics[width=0.43\textwidth]{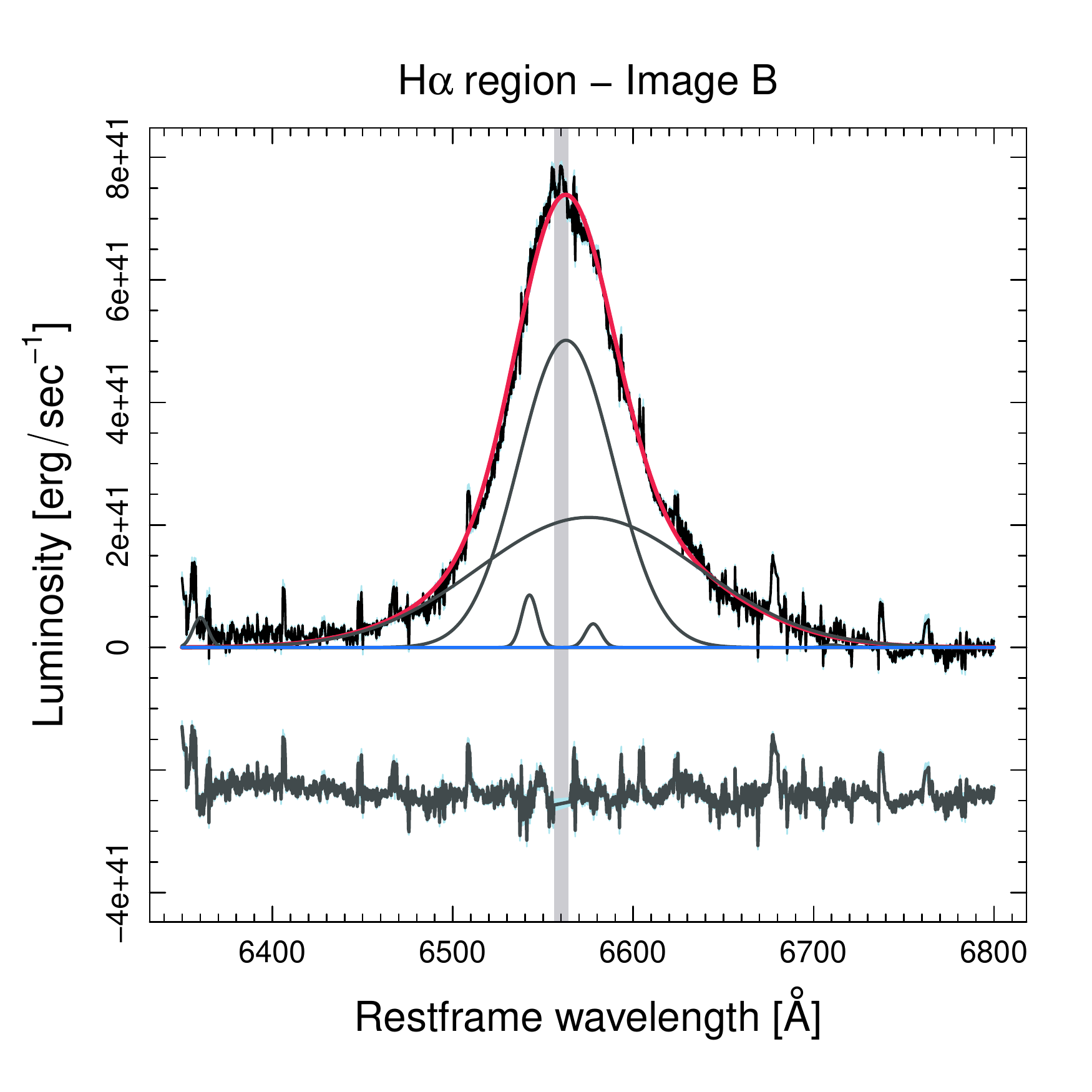}
    \end{subfigure}
        \hfill
    \begin{subfigure}{1.0\textwidth}
  \includegraphics[width=0.43\textwidth]{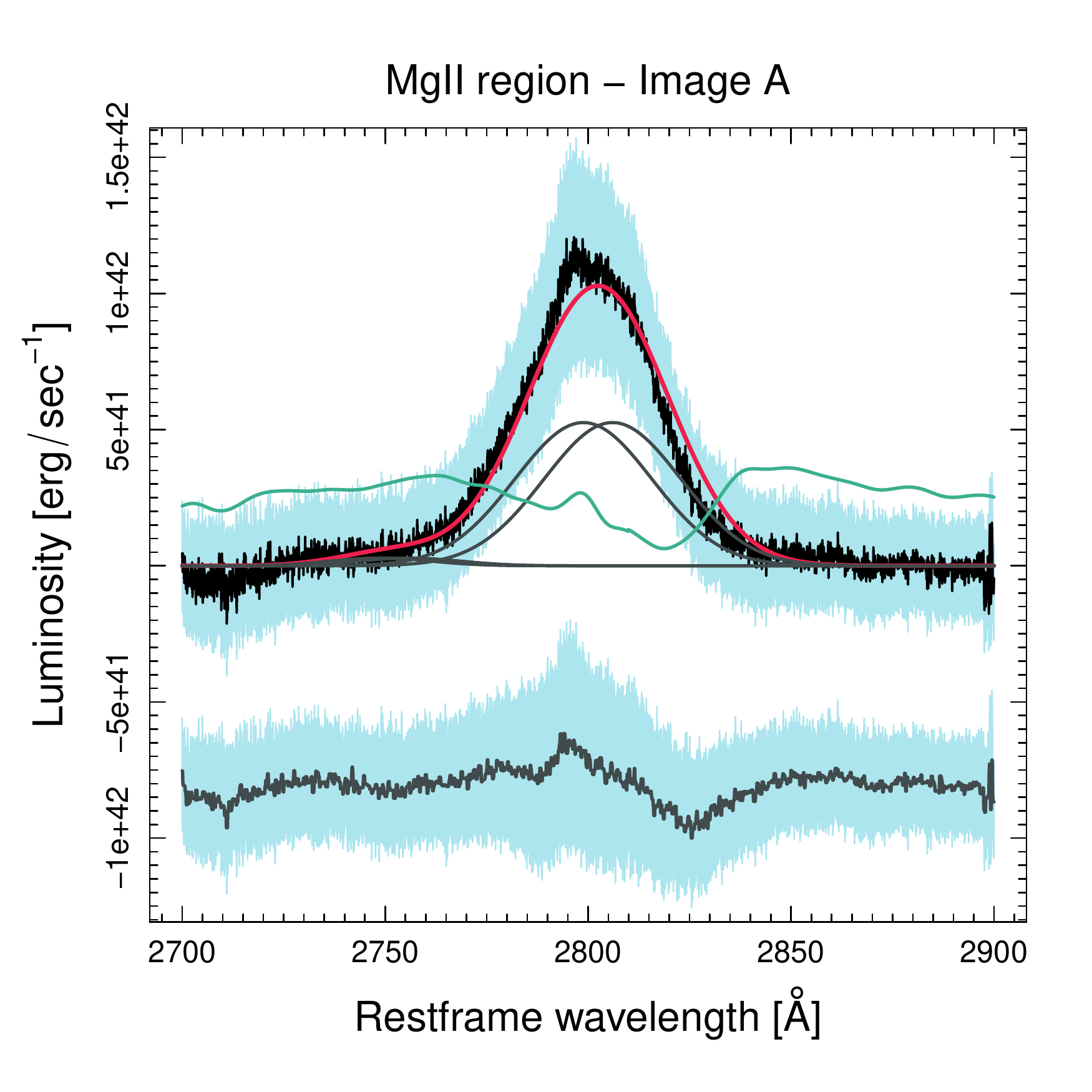}
  \includegraphics[width=0.43\textwidth]{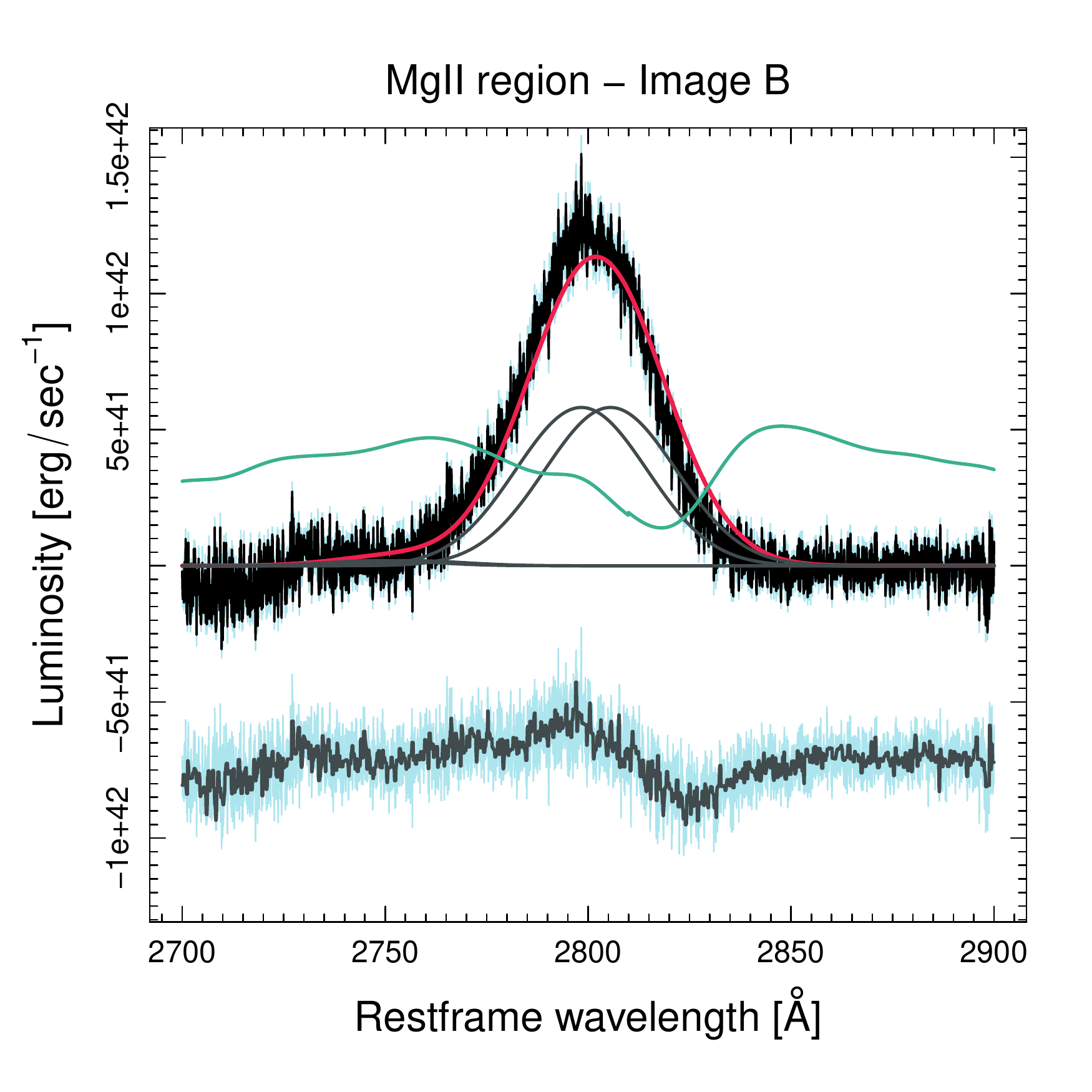}
    \end{subfigure}
    \hfill
    \begin{subfigure}{1.0\textwidth}
  \includegraphics[width=0.43\textwidth]{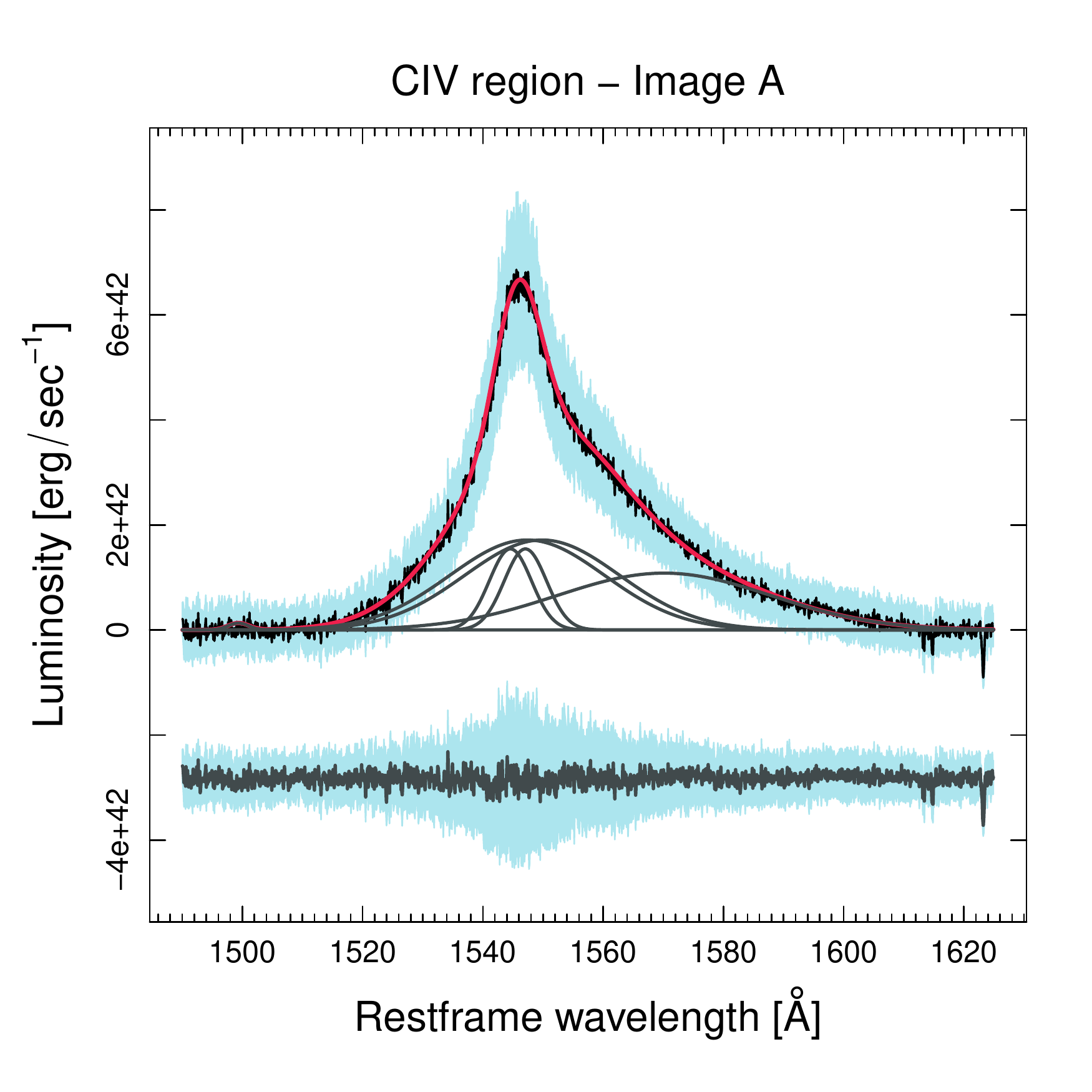}
  \includegraphics[width=0.43\textwidth]{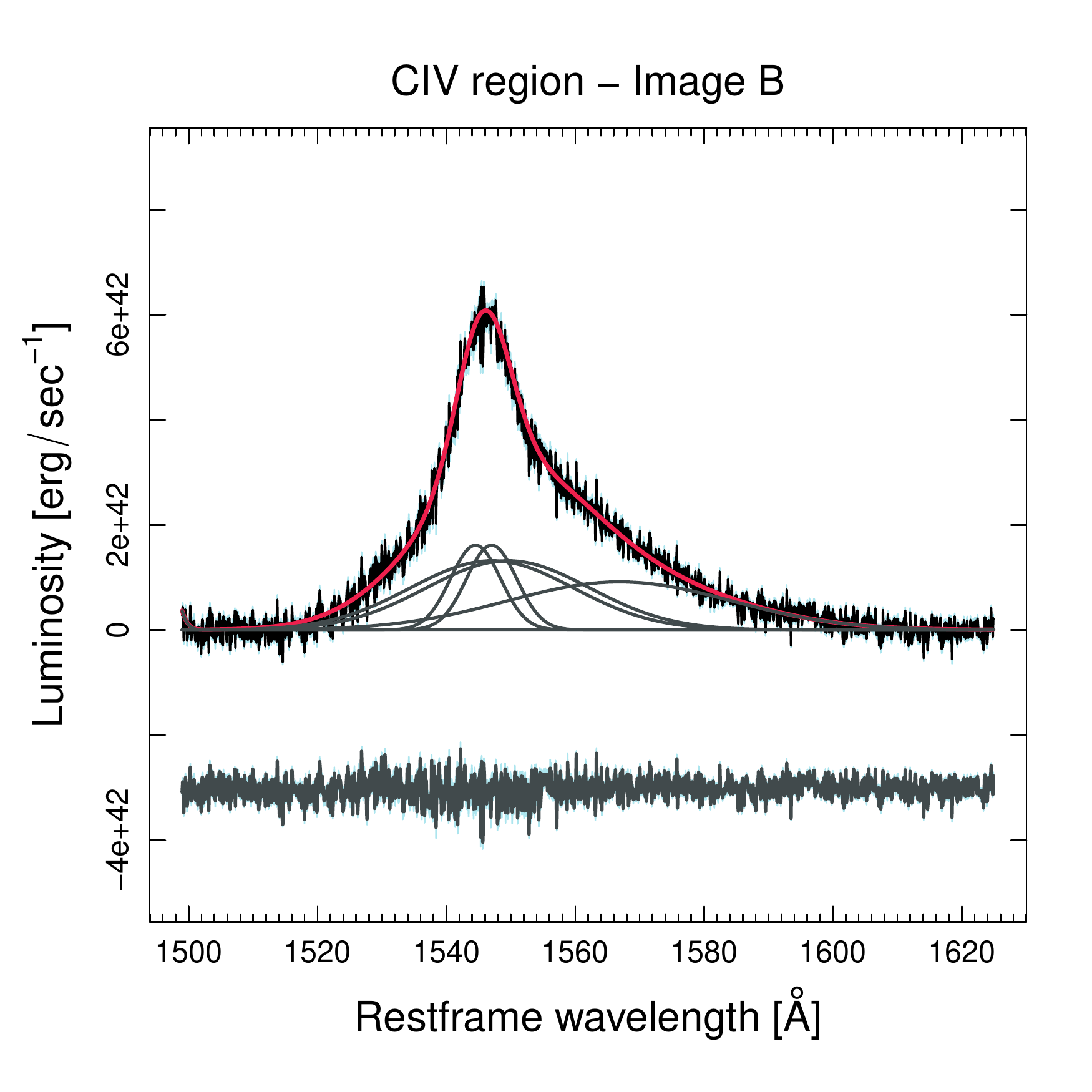}
    \end{subfigure}
      \caption{Gaussian fits to the A and B image broad emission lines of Q1355-2257. The red line is the best fit, the black lines are the different components of each region (emission and absorption), the green line is the Fe template and the blue line is the continuum fit. The 1-sigma errors are shown by the blue regions and the model residuals are shown below each spectrum.}
    \label{FigmbhQ1355}%
\end{figure*}


\begin{figure*}[!htbp]
    \begin{subfigure}{1.0\textwidth}
  \includegraphics[width=0.44\textwidth]{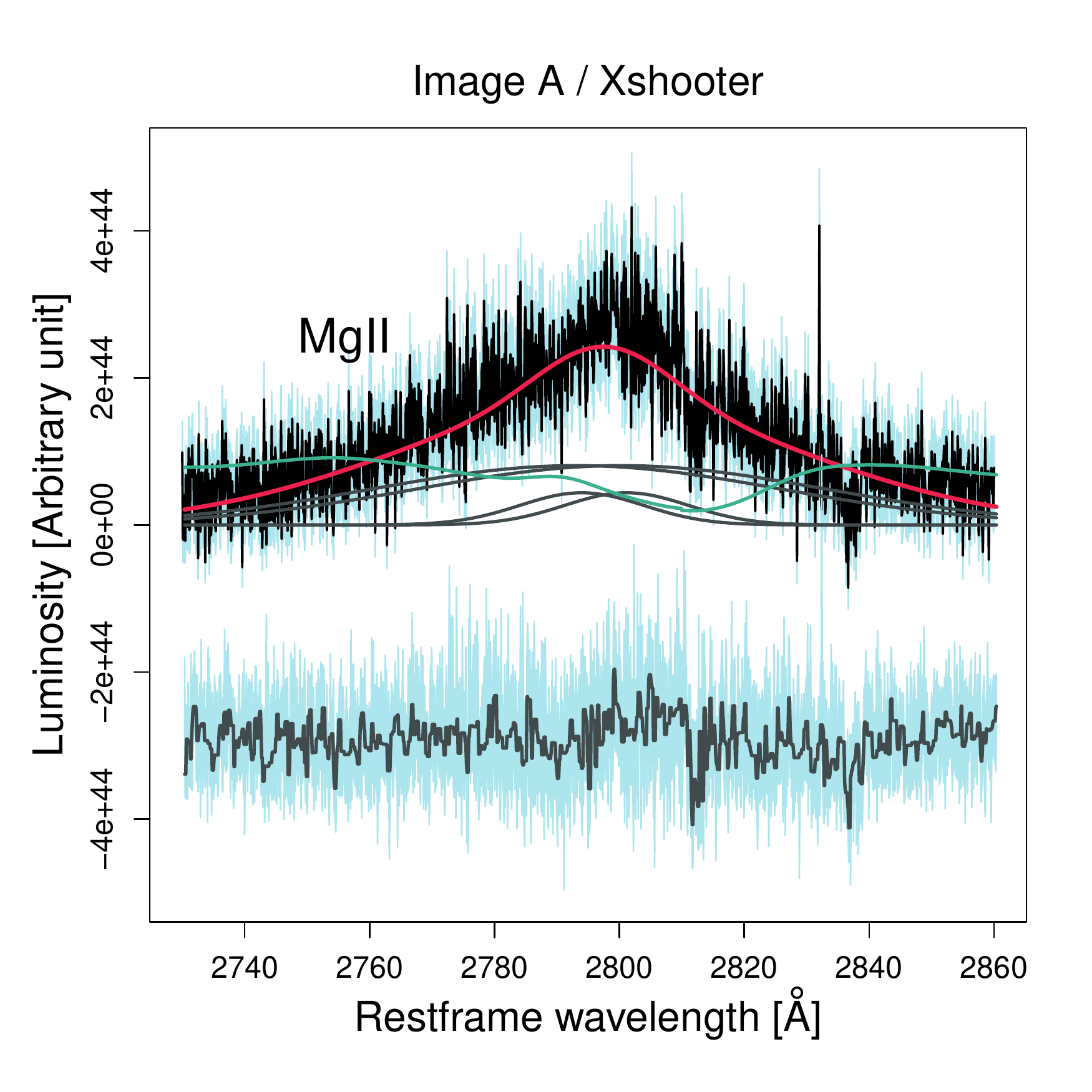}
   \includegraphics[width=0.44\textwidth]{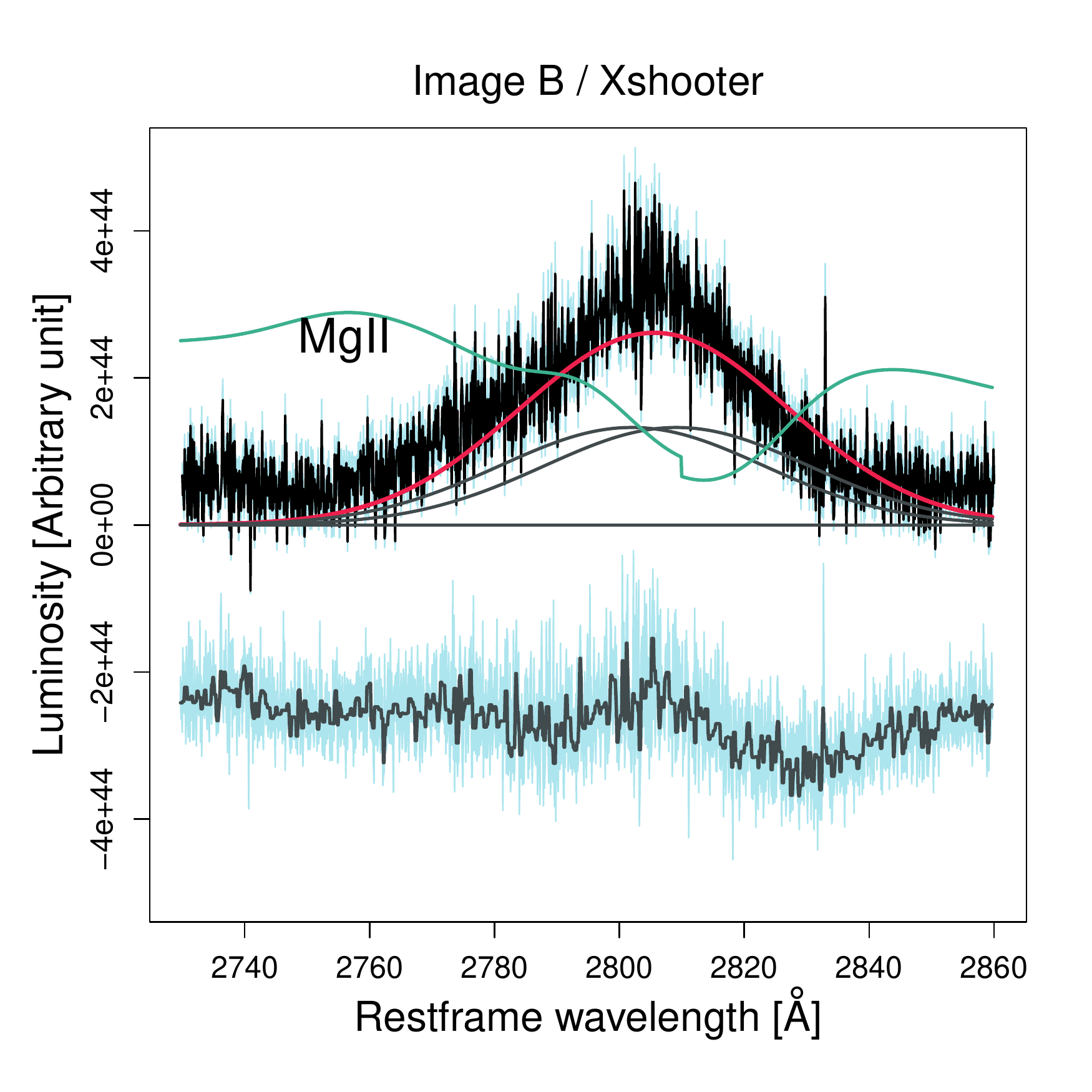}
    \end{subfigure}
      \caption{Gaussian fits to the A and B image broad emission lines of SDSS1226$-$0006. The red line is the best fit, the black lines are the different components of each region (emission and absorption), the green line is the Fe template and the blue line is the continuum fit. The 1-sigma errors are shown by the blue regions and the model residuals are shown below each spectrum.}
    \label{Figmbh1226}%
\end{figure*}

\subsection{Luminosity measurements}\label{sec:lum}

We follow \citet{2011assef} and estimate the monochromatic luminosity of each quasar using the broad band spectral energy distribution (SED) of the brightest image (A) using the fluxes from CASTLES and other sources in the literature (see Table~\ref{tab:photometric_data}). This method was preferred over using the continuum obtained from the spectra due to several factors affecting the LUCIFER and MMIRS data (e.g. low S/N (3-18), unresolved images in the slit, seeing conditions varying between the target and the standard star) and because of the chromatic microlensing detected in the continuum of the four systems observed with X-shooter and FORS2 (Melo et al. in prep.). To demagnify the fluxes, we use the magnification estimated from a lens model (Table~\ref{table:maglensmodel}). We chose photometric data that were obtained close in time to our observations to minimize differences in the amount of microlensing or a large intrinsic variation that coupled with the time delay could mimic chromatic microlensing. If light curves were available, we included the variability amplitude as part of the flux uncertainties. For instance, \citet{2017Giannini} demonstrated that HE0047$-$1756 varied by $\sim$0.2-0.3 over a five-year period, and WFI2033$-$4723 varies by 0.5 mag in four years. The system HE0435$-$1223 varied $\sim$0.4 mag (\citealt{2011Ricci}) and more recently, \citet{2017bonvin} presented 13-year light curves, with a variability ambplitude of $\sim$ 0.7 mag. 

\begin{table*}[!h]
\caption{Magnification values used for demagnifying the flux and their references.}       
\label{table:maglensmodel}      
\centering          
\begin{tabular}{ l l D{.}{.}{2.2} c l }
\toprule
Object & $z_{s}$ & \multicolumn{1}{c}{Magnification} & Image & Ref.\\
\midrule
HE0047$-$1756 &  1.66  & 13.87 & A & \citet{2014Rojas} \\ 
QJ0158$-$4325 & 1.29 & 25.27 & A & \citet{2019bhatiani} \\
HE0435$-$1223 & 1.689 & 7.27 & A & \citet{2018FianHE0435} \\ 
HE0512$-$3329  & 1.57 & 7.44 & A & \citet{2009Mediavilla}  \\ 
SDSS0924$+$0219  & 1.524  & 21.05 & A & \citet{2009Mediavilla}  \\ 
Q1017$-$207 & 2.55 	& 4.54 & A & \citet{2009Mediavilla} \\ 
HE1104$-$1805  & 2.32  & 16.20 & A & \citet{2011assef} \\
SDSS1138$+$0314 & 2.44  & 7.30 & A & \citet{2006Eigenbrod} \\
SDSS1226-0006 & 1.12 & 3.83 & A & \citet{2012Sluse}\\
LBQS1333+0113 & 1.57   & 3.77 & A & \citet{2012Sluse}  \\
Q1355$-$2257 & 1.37 & 2.50 & A & \citet{2012Sluse} \\
WFI2026$-$4536 &2.23 & 14.20 & A1 & \citet{2018Bate}  \\ 
WFI2033$-$4723  & 1.66   & 3.13 & A & \citet{2012Sluse} \\
HE2149$-$2745  & 2.03   &  2.71 & A & \citet{2012Sluse}\\
\bottomrule
\end{tabular}
\end{table*}

\begin{table*}[h!]
    \centering
      \begin{small}
    \begin{tabular}{c|c|c|c|c}
    \hline\hline
      System & Instrument & Filter & Magnitude [mag] & Ref.\\
      \hline
      HE0047$-$1756 & HST\tablefootmark{1} & F160W & 15.33 $\pm$ 0.02 & CASTLES\tablefootmark{2}\\
     & HST & F555W & 17.57 $\pm$ 0.18 & CASTLES\\
     & HST & F814W & 16.86 $\pm$ 0.15 & CASTLES\\
    & GAIA  & GAIA DR1 & 16.542 $\pm$ 0.2 & GLQ Database\tablefootmark{3}\\
    \hline
    HE0435$-$1223 & HST & F160W & 17.31 $\pm$ 0.02 & CASTLES\\
    & HST & F555W & 18.58 $\pm$ 0.02 & CASTLES\\
    & HST & F814W & 17.84 $\pm$ 0.02 & CASTLES\\
    & MagIC\tablefootmark{4} & g & 19.00 & \citet{2002Wisotzki}\\
    & MagIC & r & 18.44 & \citet{2002Wisotzki}\\
    \hline
    HE0512$-$3329 & HST & F160W & 15.81 $\pm$ 0.02 & CASTLES\\
    & HST & F220W & 18.96 $\pm$ 0.11 & \citet{2011Munoz}\\
    & HST & F250W & 18.07 $\pm$ 0.23 & \citet{2011Munoz}\\
    & HST & F330W & 17.67 $\pm$ 0.13 & \citet{2011Munoz}\\
    & HST & F435W & 18.67 $\pm$ 0.03 & \citet{2011Munoz}\\
    & HST & F555W & 18.10 $\pm$ 0.05 & \citet{2011Munoz}\\
    & HST & F625W & 17.60 $\pm$ 0.05 & \citet{2011Munoz}\\
    & HST & F814W & 16.98 $\pm$ 0.03 & \citet{2011Munoz}\\
    \hline
    SDSS0924+0219 & HST & F160W & 17.96 $\pm$ 0.02 & CASTLES\\
    & HST & F555W & 19.61 $\pm$ 0.01 & CASTLES\\
    & HST & F814W & 18.75 $\pm$ 0.05 & CASTLES\\
    & SDSS\tablefootmark{5} & u &  19.66 $\pm$ 0.02 & \citet{2003Inada}\\
    & SDSS & g &  19.46 $\pm$ 0.01 & \citet{2003Inada}\\
    & SDSS & r &  18.97 $\pm$ 0.01 & \citet{2003Inada}\\
    & SDSS & i &   18.87 $\pm$ 0.02 & \citet{2003Inada}\\
    \hline
    Q1017$-$207 & HST & F160W & 15.66 $\pm$ 0.03 & CASTLES\\
    & HST & F555W & 17.43 $\pm$ 0.03 & CASTLES\\
    & HST & F814W & 16.92 $\pm$ 0.02 & CASTLES\\
    \hline
    HE1104$-$1805 & HST & F160W & 15.91 $\pm$ 0.01 & CASTLES\\
    & HST & F555W & 16.92 $\pm$ 0.06 & CASTLES\\
    & HST & F814W & 16.40 $\pm$ 0.03 & CASTLES\\
    & Spitzer & IRAC 3.6 & 14.03 $\pm$ 0.04 & \citet{2007Poindexter}\\
    & Spitzer & IRAC 4.5 & 13.285 $\pm$ 0.07 & \citet{2007Poindexter}\\
    & Spitzer & IRAC 5.8 & 12.195 $\pm$ 0.03 & \citet{2007Poindexter}\\
    & Spitzer & IRAC 8.0 & 10.87 $\pm$ 0.03 & \citet{2007Poindexter}\\
    \hline
    SDSS1226$-$0006 & HST & F160W & 17.24 $\pm$ 0.02 & CASTLES\\
    & HST & F555W & 18.57 $\pm$ 0.01 & CASTLES\\
    & HST & F814W & 18.84 $\pm$ 0.10 & CASTLES\\
    & SDSS & i & 18.23 & \citet{2008Inada}\\
     \hline
    SDSS1138+0314 & HST & F160W & 17.96 $\pm$ 0.02 & CASTLES\\
    & HST & F555W & 19.37 $\pm$ 0.07 & CASTLES\\
    & HST & F814W & 19.05 $\pm$ 0.01 & CASTLES\\
    & GAIA & GAIA DR2 & 19.683 & GQL Database\\
    \hline
    LBQS1333+0113 & HST & F160W & 16.18 $\pm$ 0.02 & CASTLES\\
    & SDSS & u & 18.54 & \citet{2004Oguri}\\
    & SDSS & g & 18.12 & \citet{2004Oguri}\\
    & SDSS & r & 17.95 & \citet{2004Oguri}\\
    & SDSS & i & 17.60 & \citet{2004Oguri}\\
    & SDSS & z & 17.49 & \citet{2004Oguri}\\
    \hline
    Q1355$-$2257 & HST & F160W & 15.91 $\pm$ 0.02 & CASTLES\\
    & HST & F555W & 17.61 $\pm$ 0.12 & CASTLES\\
    & HST & F814W & 17.21 $\pm$ 0.07 & CASTLES\\
    & MagIC & g & 17.707 & \citet{2003Morgan}\\
    & MagIC & r & 17.322 & \citet{2003Morgan}\\
    & MagIC & i & 17.338 & \citet{2003Morgan}\\
    & MagIC & z & 17.427 & \citet{2003Morgan}\\
    \hline
    WFI2026$-$4536 & HST & F160W & 15.64 $\pm$ 0.01 & CASTLES\\
    & MagIC & i & 17.109 & \citet{2004Morgan}\\
    & PANIC\tablefootmark{6} & Ks & 14.978 & \citet{2004Morgan}\\
    \hline
    WFI2033$-$4723 & HST & F160W & 17.22 $\pm$ 0.02 & CASTLES\\
    & HST & F555W & 19.24 $\pm$ 0.03 & CASTLES\\
    & HST & F814W & 18.15 $\pm$ 0.05 & CASTLES\\
    & MagIC & i & 18.68 $\pm$ 0.01 & \citet{2004Morgan}\\
    \hline
    HE2149$-$2745 & HST & F160W & 15.67 $\pm$ 0.03 & CASTLES\\
    & HST & F555W & 16.97 $\pm$ 0.03 & CASTLES\\
    & HST & F814W & 16.52 $\pm$ 0.01 & CASTLES\\
    & GAIA & GAIA DR2 & 17.003 & GQL Database\\
    \hline
    \end{tabular}
    \end{small}
    \caption{Magnitudes for image A of each system used for constructing the SED.}
    \label{tab:photometric_data}
    \tablefoot{
    \tablefoottext{1}{Hubble Space Telescope;}
    \tablefoottext{2}{\citealt{2001castle}, \url{https://lweb.cfa.harvard.edu/castles/};}
    \tablefoottext{3}{Gravitationally Lensed Quasar Database, https://research.ast.cam.ac.uk/lensedquasars/index.html;}
    \tablefoottext{4}{Magellan Instant Camera at Las Campanas Observatory.}
    \tablefoottext{5}{Sloan Digital Sky Survey.}
    \tablefoottext{6}{Persson’s Auxiliary Nasmyth Infrared Camera at the Magellan Baade telescope.}
    }
\end{table*}

\begin{table*}[h!]
\caption{H$\alpha$ and H$\beta$ Mass estimates of the observed images. }        
\label{table:table6mbh}      
\centering   
\sisetup{separate-uncertainty}
\begin{tabular}{ l l c c c c D{.}{.}{2.2} }     
\toprule    
\toprule      
Image & Line  & FWHM & log$_{10}$(L$_{ref}$)\tablefootmark{a}
& log$_{10}$( M$_{BH}$)  & log$_{10}$( r$_s$)\tablefootmark{b}  & S/N \\
& & [km/s]  & [erg/s] & [M$_\odot$] &  [cm] & \\
\midrule
\multicolumn{7}{l}{MMIRS and LUCIFER}\\
\cmidrule(l){1-2}
HE0047$-$1756 & H$\alpha$ & 2678 $\pm$ 37 &
                          44.92 $\pm$ 0.56 &
                          8.29 $\pm$ 0.21  &
                      15.44  $\pm$ 0.85     &
                         28.11   \\
HE0047$-$1756 & H$\beta$ & 2719 $\pm$ 317  &
                          44.92 $\pm$ 0.56  & 
                         8.20 $\pm$ 0.50  &
                         15.21 $\pm$ 0.47 &
                         5.9 \\
HE0435$-$1223 & H$\alpha$ & 3216 $\pm$ 579 &
                          44.77 $\pm$ 0.01 &
                         8.36 $\pm$ 0.57 &  15.49 $\pm$ 0.42 &
                         3.7
                        \\
HE0512$-$3329 & H$\alpha$ & 2629 $\pm$ 10 &
                        44.71 $\pm$ 0.92   &
                         8.14 $\pm$ 0.25   &
                         15.34 $\pm$ 0.77   &
                          12.6  \\
SDSS0924+0219 & H$\alpha$ & 2127 $\pm$ 161 &
                            44.02 $\pm$ 0.29 &
                         7.51 $\pm$ 0.50 & 14.92 $\pm$ 1.47
                         & 5.9
                         \\
SDSS0924+0219  & H$\beta$ & 1990 $\pm$ 210 &
                         44.02 $\pm$ 0.29    &
                          7.35 $\pm$ 0.10  &
                         14.64 $\pm$ 2.17  &
                          3.9 \\ 
Q1017$-$207  & H$\alpha$ &  6177 $\pm$ 925  &
                         45.74 $\pm$ 0.44   &
                         9.55 $\pm$ 1.18   &
                        16.28 $\pm$ 0.10   &
                          5.5 \\ 
HE1104$-$1805 & H$\alpha$ & 3972 $\pm$ 226 &  
                           45.28 $\pm$ 0.73 &
                            8.87 $\pm$ 0.70  &
                           15.83 $\pm$ 0.33  &
                           18.5  \\                   
SDSS1138$+$0314 & H$\alpha$& 2330 $\pm$ 138 &  
                           44.57 $\pm$ 0.31  &
                           7.95 $\pm$ 0.50  &
                          15.21 $\pm$ 1.47  &
                           10.8 \\
LBQS1333+0113 & H$\alpha$ & 4337 $\pm$ 140 &
45.48 $\pm$ 0.48 &
9.08 $\pm$ 0.60 &
15.97 $\pm$ 0.39 &
5.72\\
WFI2026$-$4536 & H$\alpha$ & 2344 $\pm$ 15 & 
                           45.07 $\pm$ 0.53  & 
                           8.28 $\pm$ 0.25   &
                        15.43 $\pm$ 0.77     &
                          6.9   \\
WFI2026$-$4536  & H$\beta$ & 1588 $\pm$ 168 &
                           45.07 $\pm$ 0.53   &
                           7.83 $\pm$ 0.35  & 
                       14.96 $\pm$ 0.63  &
                        4.9 \\
WFI2033$-$4723 &   H$\alpha$ & 2684 $\pm$ 254 &
                            44.82 $\pm$ 0.15  &
                            8.23 $\pm$ 0.23  &
                          15.40 $\pm$ 0.81  &
                           4.29 \\
HE2149$-$2745 & H$\alpha$ &  4205 $\pm$ 272 &  
                          45.88 $\pm$ 0.59   &
                           9.31 $\pm$ 0.93  &
                           16.12 $\pm$ 0.20 &
                          5.55  \\
\midrule
\multicolumn{7}{l}{X-Shooter}\\
\cmidrule(l){1-2}
\multicolumn{7}{l}{QJ0158$-$4325}\\
\cmidrule(l){1-3}
A  &  C{\footnotesize{IV}}        &  
 4880.26 $\pm$ 166.63 &
 44.49 $\pm$ 0.92 & 
8.02 $\pm$ 0.21 & 
14.42 $\pm$ 0.85 &
    11.06 \\
 &  Mg{\footnotesize{II}}  &  
 4069.70 $\pm$ 92.59 & 
 44.59 $\pm$ 0.45 &
8.50 $\pm$ 0.11 &
15.09 $\pm$ 1.13 &
  18.74 \\
  & H$\alpha$  &  
4865.23 $\pm$ 129.42 & 
44.20 $\pm$ 0.46 &
8.34 $\pm$ 0.22 &
15.47 $\pm$ 0.83 &
  9.63 \\
B  &  C{\footnotesize{IV}}        &  
 5164.00 $\pm$ 334.75 &
 44.49 $\pm$ 0.92 &
 8.07 $\pm$ 0.54 &
14.46 $\pm$ 0.44 &
    2.63  \\
 &  Mg{\footnotesize{II}}  &  
 4204.90 $\pm$ 204.77  & 
44.59 $\pm$ 0.45  &
8.53 $\pm$ 0.23 &
15.11 $\pm$ 0.81 &
  7.44 \\
   & H$\alpha$  &  
 4651.04 $\pm$ 232.36  & 
 44.20 $\pm$ 0.46 &
8.31 $\pm$ 0.41 &
15.45 $\pm$ 0.56 &
  5.80 \\
 \midrule
\multicolumn{7}{l}{SDSS1226$-$0006}\\
\cmidrule(l){1-3}  
  A & Mg{\footnotesize{II}} & 5337.24 $\pm$ 205.29 & 44.94 $\pm$ 0.54 & 8.95 $\pm$ 0.34 & 15.39 $\pm$ 0.64 & 6.8\\
  B & Mg{\footnotesize{II}} & 5331.44 $\pm$ 133.12 & 44.94 $\pm$ 0.54 & 8.95 $\pm$ 0.36 & 15.39 $\pm$ 0.62 & 4.00\\
\midrule
\multicolumn{7}{l}{LBQS1333+0113}\\
\cmidrule(l){1-3}
 A &  Mg{\footnotesize{II}}  &  
4521.71 $\pm$ 69.38 & 
45.88 $\pm$ 0.49  &
9.38 $\pm$ 0.15 &
15.67 $\pm$ 1.00  &
  15.80 \\
  & H$\alpha$  &  
4608.55 $\pm$ 69.73 & 
45.48 $\pm$ 0.48 &
9.13 $\pm$ 0.54 &
16.00 $\pm$ 0.44  &
  8.47 \\
B &  Mg{\footnotesize{II}}  &  
4508.73 $\pm$ 29.97   & 
45.88 $\pm$  0.49 &
9.37 $\pm$ 0.47 &
15.67 $\pm$ 0.50  &
  8.79 \\
   & H$\alpha$  &  
4754.73 $\pm$ 23.66   & 
45.48 $\pm$ 0.48  &
9.16 $\pm$ 0.48 &
16.02 $\pm$ 0.49 &
  8.56 \\
 \midrule
\multicolumn{7}{l}{Q1355$-$2257}\\
\cmidrule(l){1-3}
A  &  C{\footnotesize{IV}}        &  
2939.65 $\pm$ 254.82  &
45.68 $\pm$ 0.95  & 
8.29 $\pm$ 0.18 & 
14.60 $\pm$ 0.92 &
    9.69 \\
 &  Mg{\footnotesize{II}}  &  
4254.30 $\pm$ 74.41  & 
45.78 $\pm$ 0.88  &
9.26 $\pm$ 0.13 &
15.59 $\pm$ 1.06  &
  23.34 \\
  & H$\alpha$  &  
3620.07 $\pm$ 65.09 & 
45.39 $\pm$ 0.88 &
8.86 $\pm$ 0.70 &
15.82 $\pm$ 0.33  &
  23.84 \\
B  &  C{\footnotesize{IV}}        &  
2702.83 $\pm$ 120.74  &
45.68 $\pm$ 0.95  &
8.22 $\pm$ 0.16 &
14.56 $\pm$ 0.97 &
    3.41 \\
 &  Mg{\footnotesize{II}}  &  
4118.03 $\pm$ 106.32  & 
45.78 $\pm$ 0.88 &
9.23 $\pm$ 0.10  &
15.57 $\pm$ 1.17 &
  10.40 \\
   & H$\alpha$  &  
3442.36 $\pm$ 108.17 & 
45.39 $\pm$ 0.88 &
8.82 $\pm$ 0.72 &
15.79 $\pm$ 0.31  &
  14.14 \\
\midrule
\multicolumn{7}{l}{FORS2}\\
\cmidrule(l){1-2}
\multicolumn{7}{l}{SDSS1226$-$0006}\\
\cmidrule(l){1-3}
A &  Mg{\footnotesize{II}} & 4760.49 $\pm$ 295.07 & 44.94 $\pm$ 0.54 & 8.85 $\pm$ 0.45 & 15.32 $\pm$ 0.52 & 8.8\\
B &  Mg{\footnotesize{II}} & 4838.84 $\pm$ 316.89 & 44.94 $\pm$ 0.54 & 8.86 $\pm$ 0.49 & 15.33 $\pm$ 0.48 & 10.17\\
\bottomrule
\end{tabular}
\end{table*}

\subsection{Uncertainties}

We need to consider multiple factors that could contribute to the uncertainties in M$_{BH}$. For example, the BEL of one of the images could be microlensed (e.g., microlensing affecting the red wing of the H$\alpha$ emission line in HE0435$-$1223, \citealt{2014Braibant}, and the blue wing of Mg{\footnotesize{II}} for the same system in \citealt{2018fian}), leading to a larger FWHM. \citet{2021Melo} showed that even if we have a FWHM difference between the images of $>5$ sigma, the impact on M$_{BH}$ is negligible compared with other sources of errors (see below for an specific example).

\begin{figure}[h!]
    \centering
    \includegraphics[width=8cm]{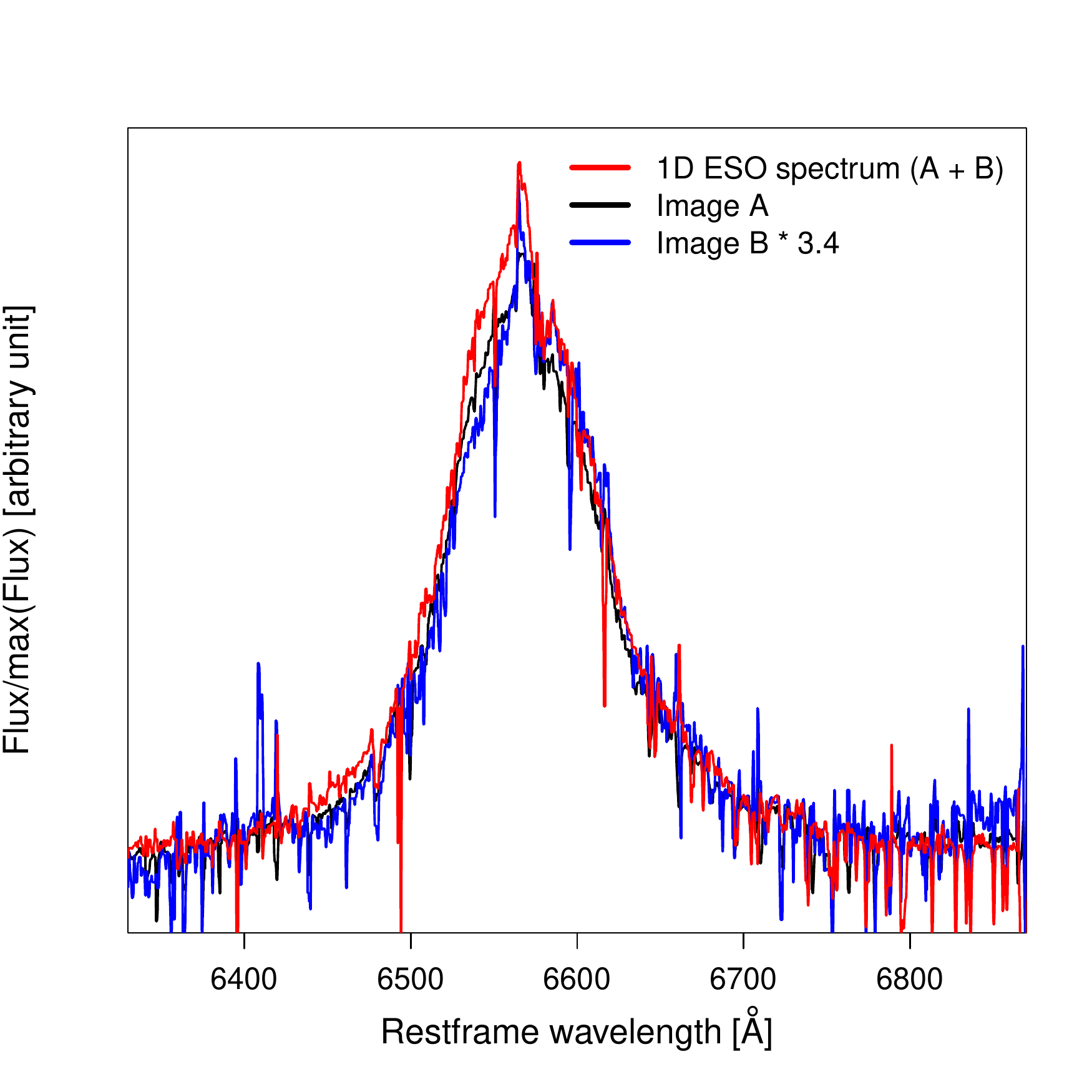}
    \caption{Combined spectra of images A+B (red spectra), compared to image A (black spectra) and image B (blue spectra) for the system LBQS1333+0113. We subtracted the continuum for the three spectra and multiplied image B by a factor of 3.4 for a clearer comparison.}
    \label{fig:lbqscomparisonfwhm}
\end{figure}

Another contribution to the uncertainties is the blending of the images in some of the MMIRS spectra. To see how much this could affect the M$_{BH}$, we compare the FWHM we find from fitting the blended image A+B spectrum of LBQS1333+0113 as compared to the separate spectra of the two images (see Fig.~\ref{fig:lbqscomparisonfwhm}).
For H$\alpha$ the FWHM of the combined spectrum is 4746.39 $\pm$ 109.89 km/s compared to 4608.55 $\pm$ 69.73 km/s for image A and 4754.73 $\pm$ 23.66 km/s for image B. These differences translate in estimated masses of log$_{10}$( M$_{BH}$/M$_{\odot}$) = 9.16 $\pm$ 0.59, 9.13 $\pm$ 0.54, and 9.16 $\pm$ 0.48 which are much smaller than the other sources of error and thus unimportant for the BH mass estimate. A similar result is obtained for the Mg{\footnotesize{II}} line.\\
Another factor contributing to the error is the monochromatic luminosity uncertainty. This has several systematic uncertainties: the systematic errors of the instrument, the magnification of the image given by the lens model, the flux calibration and intrinsic variability. To account for the intrinsic AGN variability, we add the observed variability as a contribution to the error in the monochromatic luminosity (section~\ref{sec:lum}). Although the uncertainties in the luminosity are large, the M$_{\rm BH}$ estimate scales as $L^{1/2}$, making it less sensitive to these errors compared to the FWHM because the $\rm M_{BH} \propto FWHM^2$ is so much stronger.

\section{Results}\label{Results} 

\begin{figure}[h!]
    \centering
    \includegraphics[width=8cm]{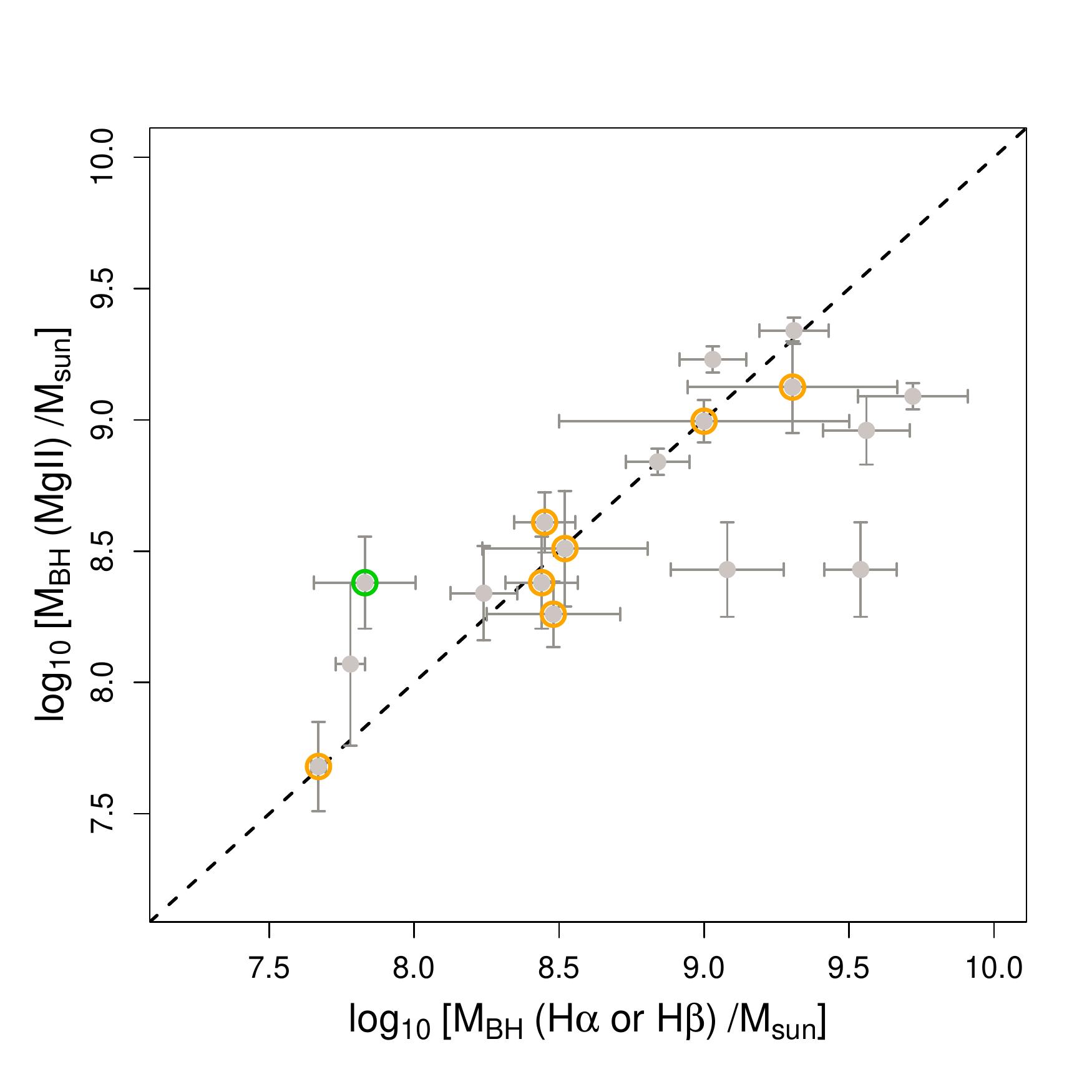}
    \caption{Comparison between M$_{BH}$ estimates obtained from the Balmer lines and Mg{\footnotesize{II}} emission lines. The new measurements are marked in orange (H$\alpha$ emission line) and green (H$\beta$ emission line). The systematic offset from \citet{2016Mejiarestrepo} is applied. The dotted line shows where the masses are equal.}
    \label{fig:halphavsmgii}
\end{figure}

Using the FWHM from the models of the emission lines and the monochromatic luminosity obtained from the SEDs, we measure M$_{\rm BH}$ following equation \ref{eq:two}. The results are shown in Table~\ref{table:table6mbh} along with their respective errors. Two systems have previous H$\alpha$ $\log_{10}(\rm M_{BH}/M_{\odot})$ (\citealt{2011assef}): HE1104$-$1805 (9.05 $\pm$ 0.23) and SDSS1138+0314 ($8.22\pm0.22$), respectively. Our estimate for HE1104$-$1805 is in agreement given its error ($8.87\pm0.70$), while for SDSS1138+0314 the result is $0.27$ dex smaller
($7.95\pm0.50$). The difference in this case is due to a combination of factors: 1) we obtain a smaller FWHM ($2330\pm138$ ~km/s versus $4700\pm200$~km/s), 2) a lower luminosity (log$_{10}$( L$_{5100}$) = 44.57 $\pm$ 0.31) vs log$_{10}$( L$_{5100}$) = 44.81, 3) , 3) low S/N ($\sim$ 10 for the spectra of image A in this work vs $\sim$ 8 for presented in \citealt{2011assef}).\\
There are the first M$_{\rm BH}$ estimate obtained for the systems QJ0158$-$4325 (log$_{10}$( M$_{\rm BH}$/M$_{\odot}$) = 8.05 $\pm$ 0.58, 8.51 $\pm$ 0.25 and 8.32 $\pm$ 0.46 for C{\footnotesize{IV}}, Mg{\footnotesize{II}} and H$\alpha$, respectively), HE0512$-$3329 (log$_{10}$( M$_{\rm BH}$/M$_{\odot}$) = 8.14 $\pm$ 0.25) and WFI2026$-$4536 (log$_{10}$( M$_{\rm BH}$/M$_{\odot}$) = 8.28 $\pm$ 0.25 and 7.83 $\pm$ 0.35, for H$\alpha$ and H$\beta$, respectively).\\
The systems HE0047$-$1756, HE0435$-$1223, SDSS0924+0219, SDSS1226$-$0006,  LBQS1333+0113, Q1355$-$2257 and WFI2033$-$4723 have previous estimates of M$_{\rm BH}$ using the Mg{\footnotesize{II}} emission lines (\citealt{2006Peng,2012Sluse, 2017DingI}) which we compare to our Balmer lines estimates in 
Figure~\ref{fig:halphavsmgii}. 
Lensed quasars that have one or both M$_{\rm BH}$ estimates presented in this work are shown in color. In general, our estimates are well correlated after we apply the offset of \citealt{2016Mejiarestrepo} (0.16 dex for M$_{BH}$ measured with H$\alpha$, and 0.25 dex using Mg{\footnotesize{II}}). The systems in which the M$_{BH}$ differ for both lines (FBQ0951+2635,B1422+231 and Q2237+030) were obtained by different authors using different methods (\citealt{2011assef,2012Sluse}) and different epochs.

\begin{figure*}[htb!]
\includegraphics[width=0.46\textwidth]{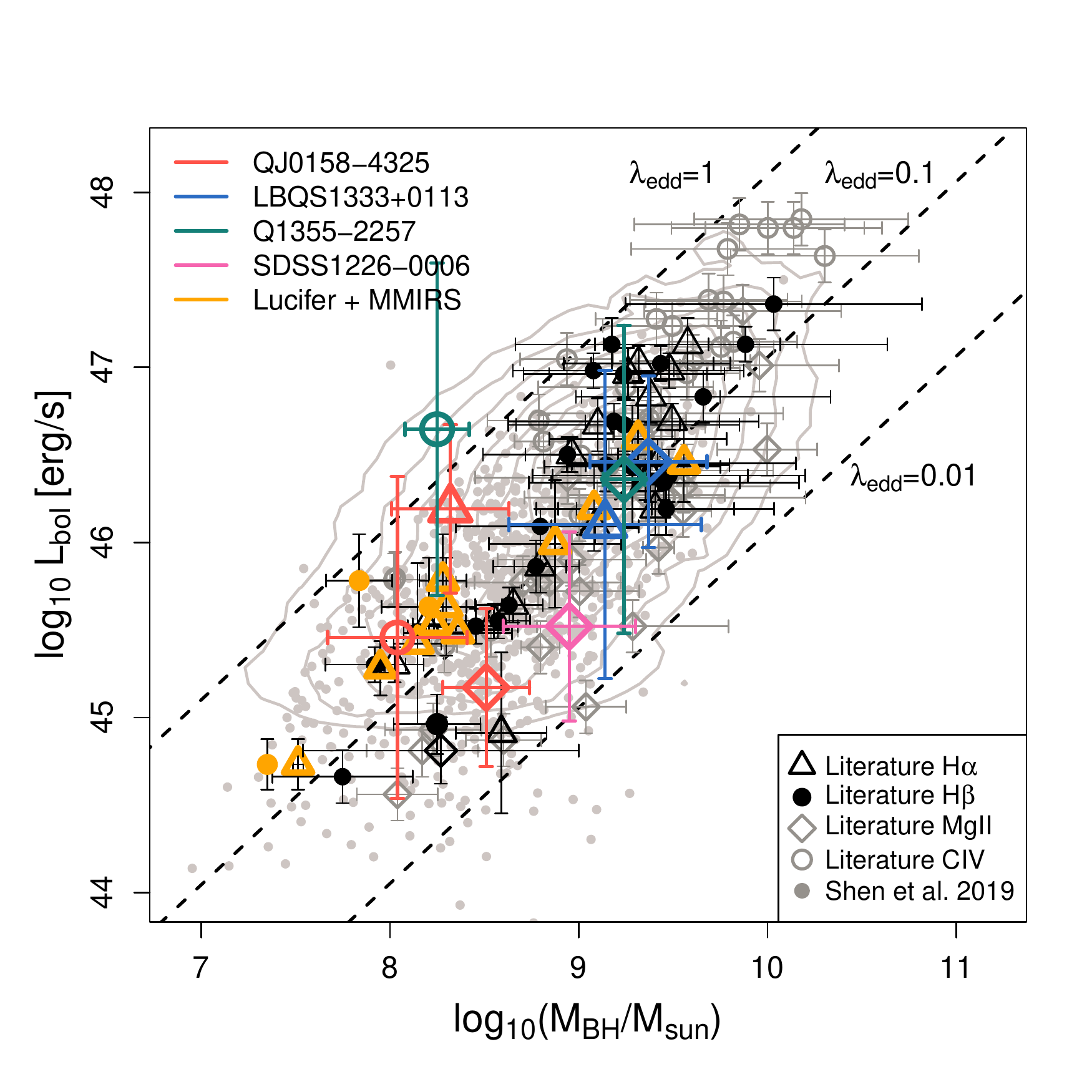}
\hspace*{0.4cm}\includegraphics[width=0.46\textwidth]{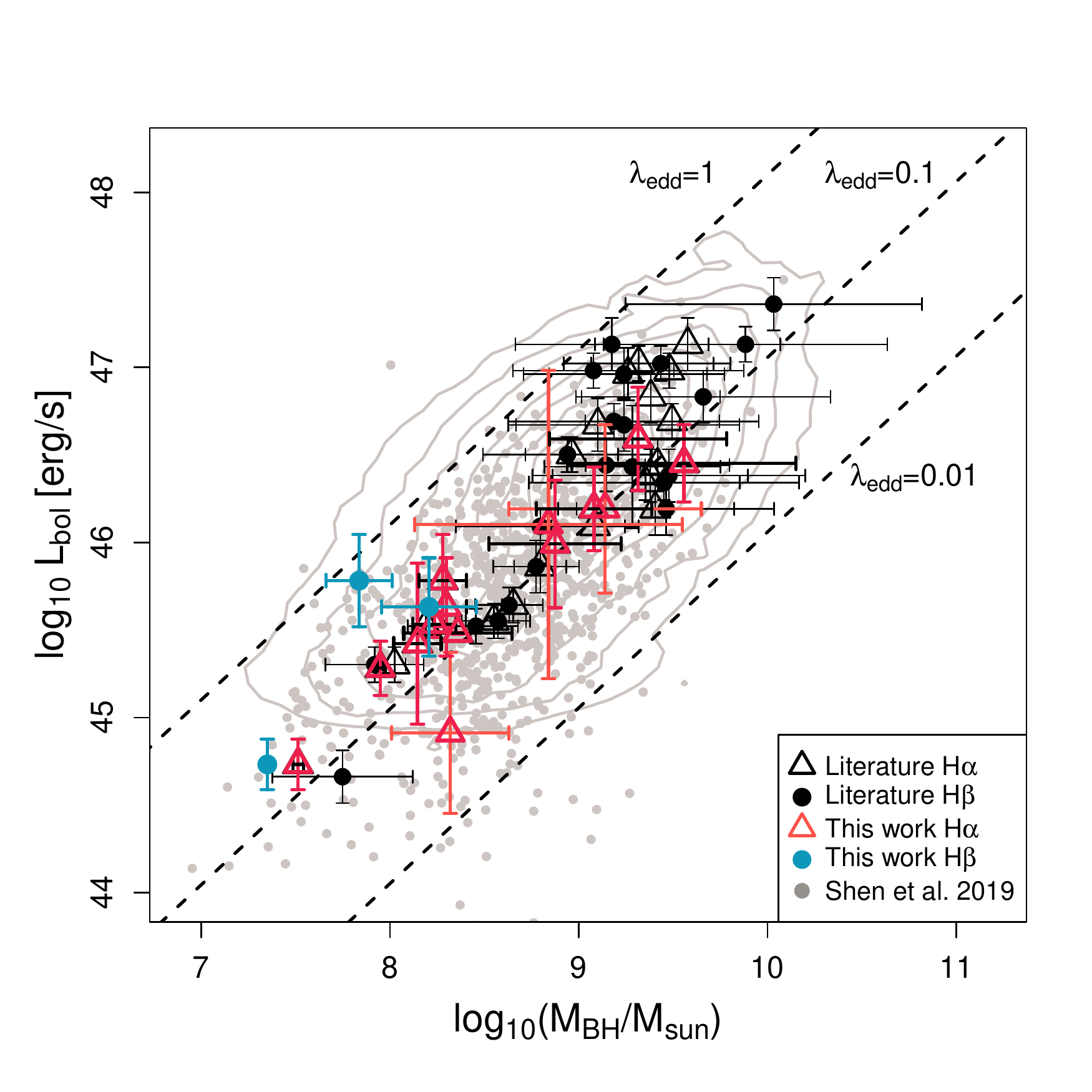}
 \caption{(Left) Logarithmic M$_{BH}$ and bolometric luminosity for all available lensed quasars (\citealt{2010greene,2012Sluse,2011assef,2006Peng,2021Melo}), and for non-lensed quasars from \citet{2011shen,2019shen}. The open triangle are the M$_{BH}$ estimates using H$\alpha$ emission line, filled circle H$\beta$, open diamond Mg{\footnotesize{II}}, and open circle C{\footnotesize{IV}}. The measurements from this study are marked with color. (Right) Logarithmic M$_{BH}$ and bolometric luminosity for just the H$\alpha$ and H$\beta$ emission lines. The red and blue points are our estimates. The dashed lines correspond to Eddington ratios of $\lambda$=1, 0.1 and 0.01.}
 \label{fig:all}
\end{figure*}

The left panel of Figure~\ref{fig:all}  shows the distribution in M$_{BH}$ and L$_{bol}$ for our systems along with estimates from the literature for 34 lensed quasars (\citealt{2006Peng,2010greene,2011assef,2012Sluse,2021Melo}). Figure~\ref{fig:all} (right) shows the distribution in luminosity and black hole mass only considering the estimates from the Balmer lines. The Eddington ratios of the lensed quasars are typically close to $\sim$ 0.1, which agrees with the results from \citet{2019shen} based on single-epoch virial BH masses of quasars. Some of the systems have several values obtained from different emission lines. The intrinsic luminosity was converted to bolometric using L$_{bol}$ = A $\times$ L$_{ref}$, where A = ( 3.81, 5.15, 9.6 ) for L$_{ref}$ = ( L$_{1350}$, L$_{3000}$ , L$_{5100}$ ) from \citet{2012Sluse}. M$_{BH}$ values obtained for non-lensed quasars using the single epoch method by \citet{2019shen} are included as the contoured distribution for comparison. In general, the new M$_{BH}$ obtained from the Balmer lines span the same range of masses as the lensed and non-lensed AGNs (Figure~\ref{fig:all}). In particular, we were able to obtain estimates for the lower luminosity systems QJ0158$-$4325, SDSS0924+0219, HE0512$-$3329 and HE0047$-$1756 (from 10$^{44}$ to 10$^{46.5}$). The systems QJ0158$-$4325 and SDSS0924+0219 have the lowest luminosities (log$_{10}$( L$_{ref}$) $<$ 44.60 L$_{\odot}$), and the latter has the lowest M$_{BH}$, log$_{10}$( M$_{BH}$/M$_{\odot}$) = 7.43 $\pm$ 0.05 (this is the average of the H$\alpha$ and H$\beta$ estimates). \\
We separately examine the three systems observed with X-Shooter (QJ0158$-$4325, LBQS13333+0113, and Q1355$-$2257) because they have multiple M$_{\rm BH}$ estimates using different emission lines. In the case of LBQS1333+0113, we only use H$\alpha$ and Mg{\footnotesize{II}} because C{\footnotesize{IV}} line exhibits multiple absorption features and H$\beta$ has low S/N (Melo et al. in prep). Mg{\footnotesize{II}} and H$\alpha$ are in good agreement with mean values of log$_{10}$(M$_{\rm BH}$/M$_{\odot}$) = 9.37 $\pm$ 0.31, and 9.14 $\pm$ 0.51, respectively. The FWHM of the H$\alpha$ emission line observed with MMIRS is in agreement given its errors with that obtained with X-shooter. \citet{2012Sluse} obtained M$_{BH}$ from the Mg{\footnotesize{II}} (log$_{10}$(M$_{BH}$/M$_{\odot}$) = 9.19 $\pm$ 0.26), which agrees with our X-shooter result.
Q1355$-$2257 exhibits a wide range of mass estimates depending on the emission line with mean values of log$_{10}$( M$_{BH}$/M$_{\odot}$) = 8.25 $\pm$ 0.17, 9.24 $\pm$ 0.11 and 8.84 $\pm$ 0.71 for C{\footnotesize{IV}}, Mg{\footnotesize{II}} and H$\alpha$, respectively (green color in figure~\ref{fig:all}). 
The Mg{\footnotesize{II}} measurement from \citet{2012Sluse} (log$_{10}$( M$_{BH}$/M$_{\odot}$) = 9.04 $\pm$ 0.34) agrees with our estimate using the same line.  As in the previous case, the C{\footnotesize{IV}} emission line for QJ0158$-$4325 is not consistent with the other estimates. The mean values for Mg{\footnotesize{II}} and H$\alpha$ are log$_{10}$( M$_{BH}$/M$_{\odot}$) = 8.51 $\pm$ 0.25 and log$_{10}$( M$_{BH}$/M$_{\odot}$) = 8.32 $\pm$ 0.46, respectively.\\
We can also estimate the unlensed size of the quasar accretion disk, r$_s$ (equation 3 of \citealt{2011mosquerakochanek}) using our M$_{BH}$ estimates and assuming a thin disk model (\citealt{1973shakurasunyaev}). The details of the parameters used are in \citet{2021Melo} and the size estimates are shown in table~\ref{table:table6mbh}. SDSS0924+0219 has the smallest accretion disk size (mean value between H$\alpha$ and H$\beta$ emission line of $r_s = 10^{14.78 \pm 2.62}$cm, an error in dex of 5.99. These spectra had very low signal-to-noise ($\sim$5.9  and $\sim$3.9 in H$\alpha$ and H$\beta$ lines, respectively).
The mean value for the systems QJ0158$-$4325, SDSS1226$-$0006, LBQS1333+0113 and Q1355$-$2257 (all emission lines from both images excluding  C{\footnotesize{IV}} are $10^{15.28 \pm 1.28}$cm, $10^{15.39 \pm 0.89}$cm and $10^{15.84 \pm 1.13}$cm, respectively. 


\section{Conclusions}\label{Conclusion}

We estimated M$_{\rm BH}$ using the broad Balmer emission lines of 14 lensed quasars measured using four different spectographs (LUCI, MMIRS, X-shooter and FORS2). After reducing and extracting the spectra corresponding to each image, the FWHM of the broad emission lines were estimated with the standard deviation of the model line profile after subtracting the narrow line components. The monochromatic luminosities were estimated using the de-magnified SED of the brightest image, taking into account the variability (if any) in the uncertainty budget.\\
These are the first M$_{BH}$ estimates for the systems QJ0158$-$4325, HE0512$-$3329 and WFI2026$-$4536. We also calculated M$_{BH}$ using the Mg{\footnotesize{II}} emission line for the systems QJ0158$-$4325, SDSS1226$-$0006, LBQS13333+0113 and Q1355$-$2257.\\
We compared the new M$_{BH}$ Balmer line to previous Mg{\footnotesize{II}} M$_{BH}$ estimates for HE0047$-$1756, HE0435$-$1223, SDSS0924+0219, SDSS1226$-$0006, LBQS1333+0113, Q1355$-$2257 and WFI2033$-$4723. The mass estimates are well correlated, with the exception of three lensed quasars (FBQ0951+2635, B1422+231 and Q2237+030) where the Balmer masses were not derived here.\\
The new Balmer M$_{BH}$ span the same range of masses estimates as non-lensed quasars with the systems QJ0158$-$4325, SDSS0924+0219, HE0512$-$3329, and HE0047$-$1756 being the lowest luminosities. The masses of the lensed quasars imply low Eddington ratios ($\sim$0.1), in agreement with the results of \citet{2019shen} from single-epoch black hole masses of SDSS quasars.

Three systems observed with X-shooter (QJ0158$-$4325,
LBQS13333+0113, and Q1355$-$2257) were analyzed in detail because they have multiple M$_{BH}$ estimates using different emission lines. 
A decade after the initial black hole mass measurements for gravitational lens systems (\citealt{2006Peng,2010greene,2011assef,2012Sluse}), this work expands the sample from 14 to 23 mass estimates.
The $M_{\rm BH}$ measurements of lensed quasars based on the Balmer lines show a lower dispersion ($RMS \sim 0.45$ dex) in M$_{\rm BH}$ at fixed bolometric luminosity, which is also true of non-lensed quasars (\citealt{2019shen}). 
Including the Mg{\footnotesize{II}} estimates increases the dispersion ($RMS \sim 0.65$ dex), confirming that the Balmer lines are more reliable. An even larger dispersion is observed too when including the Mg{\footnotesize{II}} lens M$_{BH}$ estimates from the literature. The recent discovery of new gravitational lens systems (\citealt{2023Lemon}) will allow us to explore in more detail the low-luminosity region.



\begin{acknowledgements}
We thank Kelly Denney for help with the experimental design of the LUCIFER and MMIRS observations. We thank Franz Bauer and Ezequiel Treister for carrying out the MMIRS observations. We thank Daniela Zúñiga Sacks for help with the reduction of the LUCI data. RJA  was supported by FONDECYT grant number 1231718 and by the ANID BASAL project FB210003. V.M. acknowledges support from ANID FONDECYT Regular grant number 1231418 and Centro de Astrof\'{\i}sica de Valpara\'{\i}so. N.G. acknowledges support by ANID, Millennium Science Initiative Program - NCN19\_171. This project has received funding from the European Research Council (ERC) under the European Union’s Horizon Europe research and innovation programme (ESCAPE, grant agreement No 101044152). The LBT is an international collaboration among institutions in the United States, Italy and Germany. LBT Corporation partners are: The University of Arizona on behalf of the Arizona university system; Istituto Nazionale di Astrofisica, Italy; LBT Beteiligungsgesellschaft, Germany, representing the Max-Planck Society, the Astrophysical Institute Potsdam, and Heidelberg University; The Ohio State University, and The Research Corporation, on behalf of The University of Notre Dame, University of Minnesota and University of Virginia.
\end{acknowledgements}

%
\bibliographystyle{aa} 
\bibliography{references} 
%
\end{document}